\documentclass{emulateapj}
\usepackage{amsmath}
\usepackage{graphicx}
\usepackage{amssymb}

\slugcomment{{\sc Accepted to ApJ:} April 13, 2011} 

\begin{document}

\title{Luminosity Functions and Point Source Properties from Multiple Chandra Observations of M81}

\author{
P. H. Sell\altaffilmark{1},
D. Pooley\altaffilmark{1,2},
A. Zezas\altaffilmark{3,4,5},
S. Heinz\altaffilmark{1},
J. Homan\altaffilmark{6},
W. H. G. Lewin\altaffilmark{6}
}
\altaffiltext{1}{Department of Astronomy, University of Wisconsin-Madison, Madison, WI, USA}
\altaffiltext{2}{Eureka Scientific, Oakland, CA, USA}
\altaffiltext{3}{Department of Physics, University of Crete, Heraklion, Greece}
\altaffiltext{4}{IESL, Foundation for Research and Technology Hellas, Heraklion, Greece}
\altaffiltext{5}{Harvard-Smithsonian Center for Astrophysics, Cambridge, MA, USA}
\altaffiltext{6}{MIT Kavli Institute for Astrophysics and Space Research, Cambridge, MA}

\begin{abstract}

We present an analysis of 15 \emph{Chandra} observations of the nearby spiral galaxy M81 taken over
the course of six weeks in May--July 2005.  Each observation reaches a sensitivity of
$\sim 10^{37}$~erg~s$^{-1}$.  With these observations and one previous deeper \emph{Chandra}
observation, we compile a master source list of 265 point sources, extract and fit their spectra,
and differentiate basic populations of sources through their colors.  We also carry out variability
analyses of individual point sources and of X-ray luminosity functions in multiple regions of M81 on
timescales of days, months, and years.  We find that, despite measuring significant variability in a
considerable fraction of sources, snapshot observations provide a consistent determination of the
X-ray luminosity function of M81.  We also fit the X-ray luminosity functions for multiple regions
of M81 and, using common parametrizations, compare these luminosity functions to those of two other
spiral galaxies, M31 and the Milky Way.

\end{abstract}


\section{Introduction}
\label{section:intro}

Deep X-ray observations of nearby galaxies allow for the study of their X-ray point source
populations in considerable detail.  Previous studies of a variety of different nearby galaxies have
been able to probe these sources individually by measuring their spectral characteristics and
luminosity variability over many different timescales.  Populations of sources can also be
scrutinized by the spatial and hardness distributions of sources and their X-ray luminosity function
(XLF).

\cite{fabbiano06} review the X-ray point source populations of numerous, nearby, star-forming
galaxies.  One of the most notable is M31 because of its proximity and general similarities to the
Milky Way.  Because M31 is the closest large spiral galaxy, numerous studies have analyzed the X-ray
point source populations in great detail with past and current X-ray observatories:  {\em Einstein}
\citep{trinchieri91}, {\em ROSAT} \citep{primini93,supper97}, {\em Beppo Sax} \citep{trinchieri99},
{\em Chandra} \citep{kong03}, and {\em XMM-Newton} \citep{shawgreening09}.  The two most recent
observations with {\em XMM-Newton} and {\em Chandra}, in particular, contain the best data yet, but
only in small segments of the galaxy at a time because the spatial extent of the galaxy is much
larger than the fields of view of these current X-ray telescopes.

In the last decade, the X-ray point source populations of many other nearby, star-forming galaxies
have also been scrutinized:  late-type spiral galaxies, irregular galaxies, and merging galaxies.  
Not long after the launch of the {\em Chandra} and {\em XMM-Newton} space telescopes,
\cite{soria02b}, \cite{soria02a}, and \cite{pence01} observed the X-ray point source populations of
M74, M83 and M101, respectively, all Sc-type spiral galaxies.  More recently, the ChaseM33 survey
observed the X-ray point source populations of another Sc-type galaxy, M33
\citep{plucinsky08,williams08}, and \cite{fridriksson08} completed a study of the X-ray point source
populations of a pair of galaxies, NGC 6946 (Sc-type) and NGC 4485/4490 (an irregular galaxy).  In
addition, in a series of papers, \cite{zezas02a,zezas02b,zezas06,zezas07} studied the X-ray point
source population in the unique environment of the merging Antennae galaxies.

X-ray point source populations from predominantly old stellar populations of early-type galaxies,
have also been examined.  In the S0 galaxy, NGC~1553, \cite{blanton01} found that most of the X-ray
emission is diffuse, with 49 sources comprising only 30\% of the light.  The rest of the galaxies
mentioned here are among the many optically bright elliptical galaxies associated with nearby galaxy
clusters.  NGC~4697 and NGC~4472 (M49), two large elliptical galaxies in the Virgo cluster, have
been observed with {\em Chandra}.  Detailed analysis of NGC~4697 by \cite{sarazin00} resolved almost
all of the hard emission and a large fraction of the soft emission into 90 point sources, most of
which are expected to be low-mass X-ray binaries (LMXBs).  Also, \cite{kundu02} and
\cite{maccarone03} analyzed the X-ray point source population of NGC~4472 and found 144 point
sources.  Lastly, a {\em Chandra} observation of the bright central galaxy in the Fornax cluster,
NGC~1399, revealed 214 point sources \citep{angelini01}.

From these and other studies of old stellar populations, general trends have been discerned in the
X-ray populations.  In the XLFs of X-ray point source populations of elliptical galaxies, there is
evidence for a varying break and varying slopes on either side of the break \citep[e.g.,][]{kim10}.
There is also evidence for differences between the XLFs of field and cluster binaries
\citep{fabbiano10}.  Particularly prevalent in the X-ray observations of large elliptical galaxies
is the large fraction of globular cluster sources
\cite[e.g., $\lesssim 70\%$ of the sources in NGC~1399;][]{angelini01}.  By matching {\em Chandra}
and {\em HST} observations, possible trends in the populations can be explored (e.g., metallicity).

In these many X-ray observations of early- and late-type galaxies, these authors inspect the X-ray
point sources populations from a variety of perspectives.  Hardness ratios are frequently calculated
(albeit in a number of different ways).  They are most often used to differentiate populations of
sources within each galaxy but have been also used to differentiate galactic sources from background
active galactic nuclei (AGN).  The most luminous or most unusual sources in each of the galaxies are
typically of particular interest:  ultra-luminous X-ray sources (ULXs), transient sources, or
supersoft sources (SSSs).  They are frequently examined in detail in terms of their spectral
characteristics and spectral and flux variability.

When X-ray point source populations are inspected using XLFs of different galaxy types and stellar
population ages (i.e., actively star-forming versus relatively quiescent), considerable variation
has been observed.  High-mass X-ray binary (HMXB) XLFs, which are associated with regions of current
star formation, are typically described by straight (unbroken) power laws.  For simplicity, we refer
to these as ``disk-like" XLFs in this study.  Typical cumulative slopes are $\sim -$0.7--1.0 for
moderate star formation rates (SFRs; $\lesssim 1 M_\sun$~yr$^{-1}$) and $\sim -$0.4--0.5 for higher
SFRs \citep[e.g.,][]{kilgard02, grimm02, colbert04}.  LMXB XLFs or what we refer to as
``bulge-dominated" XLFs in this study are not associated with recent star formation.  These XLFs are
typically described by broken or cutoff power laws \citep[e.g.,][]{kim04b, gilfanov04a} with signs
of a flat low-luminosity end and a break or cutoff at a few $\times \ 10^{38}$~erg~s$^{-1}$, which
is frequently attributed to the Eddington luminosity for a neutron star.  Spiral galaxies (earlier
type, in particular) show mixed XLFs with contributions from both types of XLFs, ``disk-like" and
``bulge-like".

Developing physical models to describe the populations of galactic X-ray sources, which are
the result of the evolution of their stellar populations, is a daunting task.  Some studies have
attempted to understand the star formation history of these galaxies through the use of population
synthesis modeling.  These complex models attempt to follow the stellar populations through many
evolutionary phases to the formation of the X-ray bright systems that we observe
\citep{belczynski08}.  There also exist alternate, less complex, methods such as that of
\cite{wu03} in a previous study of M81 and \cite{white98}, who construct simple birth and death rate
models for components of the stellar populations.  However, accurately constructing XLFs to low
luminosities and comparing them to models is very difficult \citep[e.g.,][]{fragos08}, and the
information gleaned from doing this (e.g., finding and interpreting breaks in XLFs) can be
ambiguous.

Some key questions concerning XLFs remain unanswered.  For example, is there a break in the XLFs of
spiral galaxies at low fluxes?  Also, essentially every study of these types of X-ray point sources
has noted a wide range of levels and timescales of both integrated flux variability and spectral
variability.  However, the XLFs of galaxies are almost always characterized by a single X-ray
snapshot of the galaxy.  A critical question can then be raised:  does the variability of the
individual sources significantly manifest itself in a variable XLF?  In other words, is the XLF
robust against the fluctuations of its constituents?  Very large changes in the XLF over the
timescales that we are probing, which are much shorter than stellar evolution timescales, should not
be observed since this would require correlated variability of many sources.  However, the stability
of XLFs has not been thoroughly investigated over the days, weeks, and years timescales.

The nearby Sb-type galaxy, M81, with multiple types of X-ray binaries, is a key object with
which to address these points and is an excellent choice for studying the X-ray point source
population with \emph{Chandra}.  Because M81 is relatively nearby at a well-determined distance
($3.63 \pm 0.34$~Mpc; \citealt{freedman94}; using the Cepheid period-luminosity relationship), we
can detect sources with faint luminosities in relatively short exposures and calculate their
luminosities accurately.  However, M81 is at a large enough distance so that almost the entire
galaxy fits within the field of view of \emph{Chandra} in one exposure.  These facts make it easy to
use \emph{Chandra} to study the variability characteristics of the X-ray point source population to
relatively low luminosities ($\gtrsim 10^{37}$~erg~s$^{-1}$) in a minimum amount of observations.

M81's X-ray point source population has been scrutinized many times in the past.  \cite{fabbiano88}
was the first to study the X-ray point source population of this galaxy with {\em Einstein}, but was
only able to detect a handful of sources.  Later on, \cite{immler01} studied this galaxy with
{\em ROSAT} and detected $\sim 5$ times as many sources to fainter luminosities.  Most recently,
\cite{tennant01} and \cite{swartz03} found 177 sources with \emph{Chandra} in a single exposure,
17--27 of which are expected to be background AGN \citep[e.g.,][]{rosati02}.  Out of these galactic
sources, some are expected to be SNRs, ULXs, SSSs, or young pulsars/pulsar wind nebulae, but most
are likely LMXBs and HMXBs.

We seek to follow-up the work of \cite{tennant01} and \cite{swartz03} and to add to the
understanding of the characteristics of the X-ray point sources and XLFs of nearby galaxies by
carrying out an observational study of the nearby galaxy M81.  We use a set of sixteen
\emph{Chandra} observations of M81 that explore variability on timescales of days, weeks, and years.
We also use these observations to make the most complete study of M81's X-ray point source
population to date.

In  \S~\ref{section:obs_red}, we lay out the observations and data reduction procedures.  Then, in
\S~\ref{section:source_lists}, we discuss the creation of our point source catalogs.  In
\S~\ref{section:HR} and \S~\ref{section:indiv_src_var}, we present hardness ratios and discuss the
individual variability of the point source population, respectively.  Following this, we construct
XLFs in \S~\ref{section:XLFs}.  Then, in \S~\ref{section:general_comps}, we compare our results with
M31 and the Milky Way.  Finally, in \S~\ref{section:summary}, we summarize our results and state our
conclusions.  For all of our analyses throughout this work, we use the energy range of 0.5--8.0 keV
(e.g., fluxes, luminosities), unless otherwise noted.

\section{Observations and Data Reduction}
\label{section:obs_red}

We present fifteen \emph{Chandra} observations of M81 (ObsIDs~5935-5949) taken specifically to
explore XLF and individual point source variability.  For our analysis, we include an additional
observation from the {\em Chandra} archive \citep[ObsID~735][]{tennant01,swartz03} to provide a
longer baseline for measuring variability (see Table~\ref{table:obsinfo}).  All sixteen observations
were taken with the Advanced CCD Imaging Spectrometer (ACIS; \citealt{garmire03}) in Timed Exposure
mode with a frame time of 3.2 s and the aimpoint on the default location of the S3 chip.  The data
were telemetered to the ground in Faint mode.

\begin{figure}[tbp]
\centering
\includegraphics[scale=0.35]{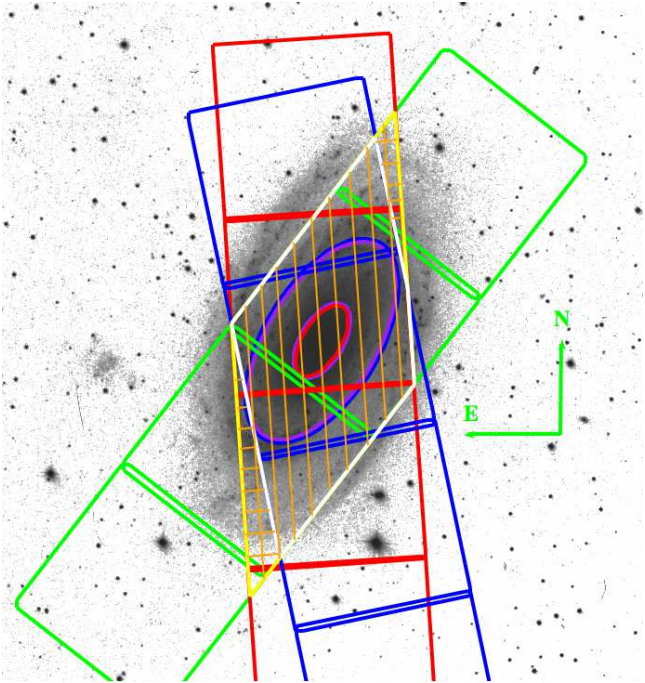}
\caption{Overlaid on a blue POSS2 DSS image of M81 are the outlines of the positions of the ACIS-S
chips to show where we have coverage in the 16 observations that we use.  The blue boxes represent
ObsID 735 and the red and green boxes represent ObsIDs 5935 and 5949 (the remainder of the 2005
observations were taken at intermediate roll angles).  The overlapping sky coverage of the fifteen
2005 and all sixteen observations are represented by the yellow and white polygons (or regions of
orange cross-hatch and vertical lines), respectively.  The ellipses mark the separations between the
bulge, inner disk, and outer disk regions (see \S~\ref{section:HR--bulge_disk_sep}; the colors of
the ellipses match the colors used for the sources in Fig.~\ref{fig:HR}).  The image is
$\sim 30^\prime \times 30^\prime$.
\label{fig:outline}}
\end{figure}

We analyze all of these observations using the CIAO software provided by the Chandra X-ray Center as
well as the IDL-based ACIS Extract program version 2008-03-04 (AE; \citealt{broos10}).  We ignore
the ACIS-I chips for the entire analysis for three primary reasons:  1) the observations were
purposely taken to line-up the galaxy on the ACIS-S chips so that any sources found on the ACIS-I
chips are far less-likely to be associated with M81; 2) point source extractions for sources on
chips I2 or I3 become very difficult because the CALDB PSF library that this version of AE uses for
automated spectral extraction does not contain the appropriate PSFs; and 3) the PSF quickly becomes
impractically large there.  Our X-ray coverage of the galaxy is shown in Fig.~\ref{fig:outline}.

\begin{deluxetable}{lllll}
\tablecaption{\emph{Chandra} Imaging Observations of M81}
\tablehead{
	\colhead{Start Date} &
	\colhead{Spacing$^\dagger$} &
	\colhead{Obs. ID} &
	\colhead{Exposure Time} \\
	& (days) & & (s) }
\startdata
2000 May  7  &	------	&	735		&	50021 \\
2005 May 26  &	1844.9	&	5935	&	10979 \\
2005 May 28  &	2.7	&	5936	&	11406 \\
2005 Jun  1  &	3.5	&	5937	&	12006 \\
2005 Jun  3  &	2.6	&	5938	&	11807 \\
2005 Jun  6  &	2.7	&	5939	&	11807 \\
2005 Jun  9  &	2.7	&	5940	&	11974 \\
2005 Jun 11  &	2.6	&	5941	&	11807 \\
2005 Jun 15  &	3.2	&	5942	&	11952 \\
2005 Jun 18  &	3.4	&	5943	&	12012 \\
2005 Jun 21  &	2.7	&	5944	&	11807 \\
2005 Jun 24  &	3.0	&	5945	&	11576 \\
2005 Jun 26  &	2.7	&	5946	&	12019 \\
2005 Jun 29  &	2.7	&	5947	&	10698 \\
2005 Jul  3  &	3.5	&	5948	&	12028 \\
2005 Jul  6  &	3.3	&	5949	&	12022 \\
\enddata
\tablecomments{($\dagger$) time since the prior observation}
\label{table:obsinfo}
\end{deluxetable}

We checked the relative pointing of \emph{Chandra} and found it to be quite good.  We only found
tiny relative, systematic offsets of not more than $0 \farcs 05$ between the observations by
comparing the positions (which were optimized with multiple iterations of AE's CHECK\_POSITIONS
stage) of all of the point sources within 2$^\prime$ of the aimpoint at one time.  We changed the
WCS header information to account for these tiny offsets.  Then, for all 16 observations, we
reprocessed the level=1 event files with CIAO 4.0 following the threads on the \emph{Chandra} X-ray
Center website\footnote{http://asc.harvard.edu/ciao/threads/createL2/} to apply the most-recent
calibration updates available (CALDB version 3.5.0).  We used all of the default
``acis\_process\_events" parameters except the ``rand\_pix\_size" keyword, which we set to 0. 
Monoenergetic exposure maps at 1.0 keV were created where needed for {\small \sc wavdetect} and the
AE analysis.

We inspected the total background for each observation and found background flares in ObsIDs 5936,
5945, 5946, and 5947.  These flares will not affect our point source extractions (although they do
affect our background model selection) because we always include a local background when fitting
each source's spectrum.  Therefore, we did not exclude the time intervals of the flares in our
observations.

\begin{figure*}
\centering
\includegraphics[scale=2.6]{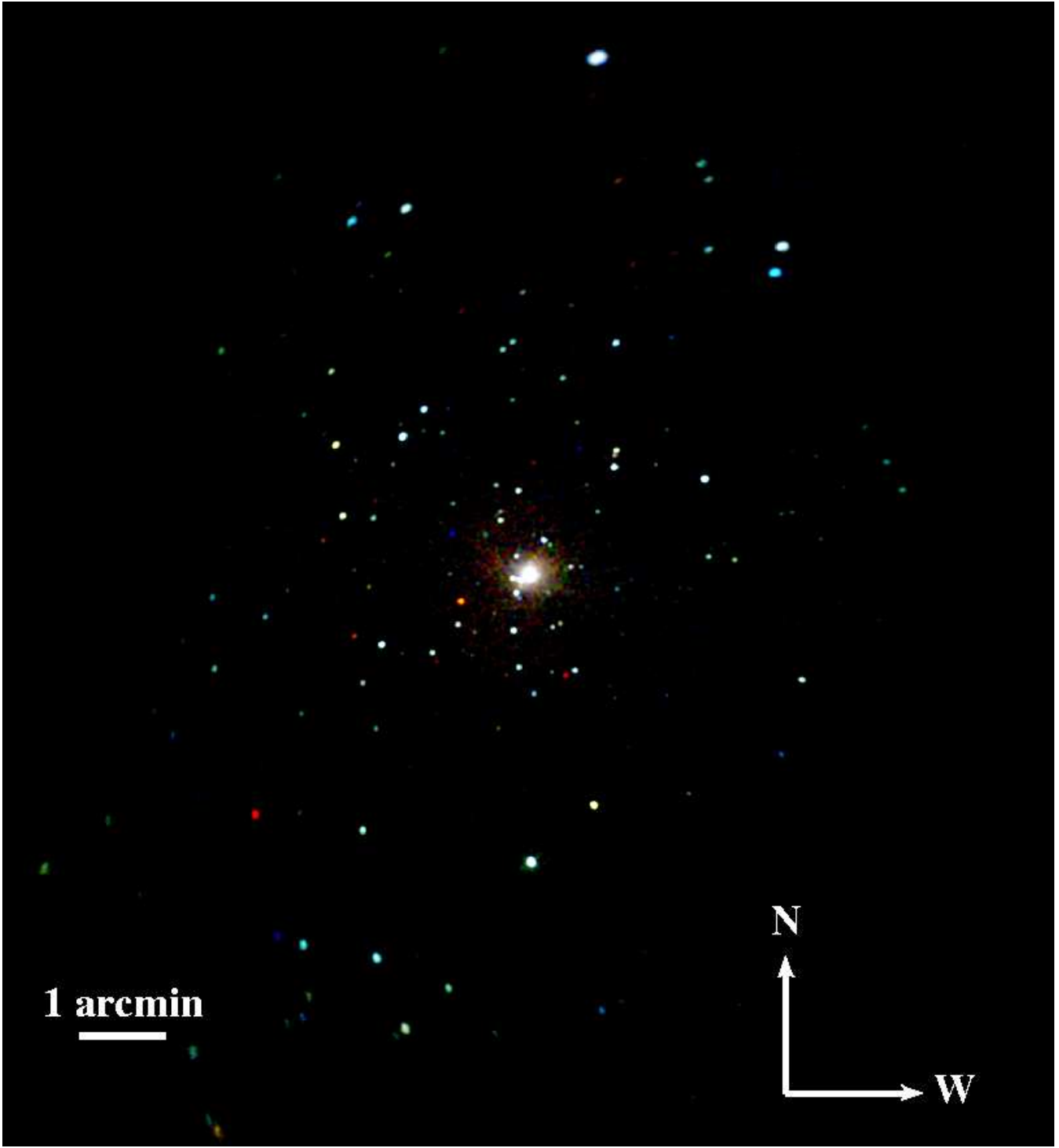}
\caption{Three-color \emph{Chandra} image of the merged observation.  Photons in the 0.5--1 keV band
(``soft'') are shown in red, those in the 1--2 keV band (``medium'') are shown in green, and those
in the 2--8 keV band (``hard'') are shown in blue.  The image is $12.4^\prime \times 14.4^\prime$.
\label{fig:3color}}
\end{figure*}

We also constructed a coadded observation of ObsIDs 5935--5949.  We used the ``merge\_all" CIAO
contributed script to coadd the event files and exposure maps for source detection and for the
creation of Figure~\ref{fig:3color}.  Any other coadded source information was calculated for each
observation with AE and then merged with AE's MERGE\_OBSERVATIONS step, which properly takes into
account observation-to-observation instrumental details (i.e., weighting the data products
appropriately).  We refer to information gleaned from the combination of ObsIDs 5935--5949 as the
``merged" observation information throughout this work.

\section{Source Lists}
\label{section:source_lists}

\subsection{Source Detection and Construction of the Master Source List}
\label{section:source_lists--master}

We used {\small \sc wavdetect} \citep{freeman02}, CIAO's wavelet source detection algorithm to
search for sources in all of the individual observations as well our merged observation.  It was not
appropriate to include ObsID 735 in the merged observation for source detection because the
aimpoint of this observation (very near to SN~1993J) is $\sim 2.8$~arcmin southwest of the aimpoint
of ObsIDs 5935--5949 (very near the center of the galaxy).  This means that the size and shape of
the PSF is different in the same sections of the galaxy between ObsIDs 5935--5949 and 735, which
would lead to less-efficient source detections.

We first filtered the ObsIDs to 0.5--6.0 keV to minimize noise.  Then, we consistently searched for
point sources on a chip-by-chip basis since the size and shape of the PSF changes as a function of
off-axis angle\footnote{the angular distance between the Chandra aimpoint and the source}.  We used
the default settings for the {\small \sc wavdetect} parameters, where applicable (notably, the
significance threshold, ``sig\_thresh", was $10^{-6}$, which corresponds to approximately one
spurious source per ACIS chip), except for the scales parameter.  We searched a variety of spatial
scales in a $\sqrt{2}$ series (as suggested in the {\small \sc wavdetect} manual) using different
upper and lower bounds for each chip.  We searched scales of 2--8 pixels on S3, 2--16 pixels on S2,
6--34 pixels on S4, and 14-40 pixels on S1.

There were a number of complications that arose when constructing our master source list:

\begin{enumerate}

\item While matching sources between observations in crowded subfields at large off-axis angles, it
was not always initially clear whether the apparently matched source was actually the same source or
whether we were confusing multiple nearby sources.  This is because the position uncertainty far off
axis can be quite large, especially when the source is near the detection limit.  Furthermore,
source variability can cause some sources to become particularly dim or drop below the detection
threshold in certain observations.  As a remedy for most of the sources in this category, we
considered two detected sources as one if their extraction regions, which enclose the inner 50\% of
the PSF, overlapped.

\item For a few sources at intermediate off-axis angles, PSF substructure also caused very close
double source detections.  Careful inspection of the PSFs for these few sources revealed that
{\small \sc wavdetect} finds two sources in one complicated PSF (see \citealt{kim04a}, who provide
details about this problem in their analysis for the ChaMP survey).  While it is possible that, in
these cases, the double sources are real, it is very unlikely and, following our rule in step 1., we
would not be able to separate them anyway if they were real.

\item A considerable readout streak from M81*, the AGN in the low-ionization nuclear emission-line
region (LINER) in the center of the galaxy, was present in all 16 observations, most significantly
ObsID~735, causing {\small \sc wavdetect} to find many false sources.  We exclude M81* from our
source lists.  Sources on the readout streaks were considered
real only if they were detected in multiple observations.  We were able to make this cut since the
read streak changed position on the sky with the changing roll angle of the telescope.

\item Obvious false detections, which include sources with only 1--2 counts were rejected.  One
false detection is expected per chip with the default {\small \sc wavdetect} settings that we used.

\item The background in the center of M81 due to the wings of the PSF of M81*, unresolved galactic
point sources, and diffuse galactic emission was non-uniform, quite high, and varied slightly in
intensity in different positions on the chips, which made it difficult to detect sources.  In this
case, we used maximum likelihood reconstructions (constructed with multiple iterations of IDL's
``max\_likelihood" routine--implemented with AE) of the region near M81*.  Six additional sources
and a severe surface brightness depression caused by pileup effects were detected in the
neighborhood of M81*.  Since these sources were embedded very near to this piled-up region and in a
highly sloped background from a combination of wings of the PSF of M81* and M81's galactic
background, we do not include these sources in our master or borderline source lists (see below).
Because of these complications, we were only able to reliably extract the {\small \sc wavdetect}
positions for these sources, which we list in Table~\ref{table:ML_sources}.

\end{enumerate}

\begin{deluxetable}{lll}
\tablecaption{Sources Near the Center of M81}
\tablehead{
	\colhead{Source} &
	\colhead{RA} &
	\colhead{Dec} \\
	\colhead{number} &
	\colhead{(degrees)} &
	\colhead{(degrees)} }
\startdata
ML1 & 148.88791 & 69.06654 \\
ML2 & 148.88738 & 69.06575 \\
ML3 & 148.89164 & 69.06485 \\
ML4 & 148.89197 & 69.06385 \\
ML5 & 148.88913 & 69.06377 \\
ML6 & 148.88531 & 69.06648 \\
\enddata
\tablecomments{Sources were found using maximum likelihood image reconstruction}
\label{table:ML_sources}
\end{deluxetable}

In any of the source rejection steps above, if we were uncertain whether to keep the source or not,
we kept it.  Out of these first five source list refinement steps, most of the sources were removed
in step 3.

\begin{enumerate}

\item[6.] We made one final cut on our preliminary source list using the AE's PROB\_NO\_SOURCE
statistic, which estimates whether a source is real by sampling the binomial probability
distribution.  We kept only sources at the 99.9\% probability level of being real according to this
statistic.  Our final source list contains 265 sources which are listed in ancillary tables.  We
also include list of sources at the 99--99.9\% probability level according to the PROB\_NO\_SOURCE
statistic and deem these 11 sources ``borderline sources" (referred to as B1-B11).  Note that only
coordinates are listed for three sources in the master source list and two sources in the borderline
source list because they were only in the field of view (on chip) of ObsID 735:  234, 241, 262, B8,
and B9.

\end{enumerate}

Overall, this careful multiple-step approach to refining our master source list rejected $\sim 36
\%$ of the original {\small \sc wavdetect} sources as false detections.

Finally, we note that using the \emph{Chandra} Point-Source Catalog (CSC) to construct source lists
for M81 for simplicity is tempting, but would yield incomplete results.  The current source list
from the CSC contains $\lesssim 50\%$ of sources that we found in the galaxy through our more
careful searching.  Such a disparate result can be explained by the differences in how we and the
CSC make use of {\small \sc wavdetect} for the construction of our source lists (our numbered
procedure above is very different than the CSC's
procedures\footnote{http://cxc.harvard.edu/csc/proc/}).  For example, the CSC uses different:
wavelet scales, energy filtering, blocking, and significance thresholds.  The most important
difference to the overall process is that we have stacked our 15 new observations, revealing a
multitude of additional, faint sources.

\subsection{Point-Source Extraction with ACIS Extract}
\label{section:source_lists--AE}

We used AE for the extraction of the source and background spectra.  One of the primary reasons that
we use AE (as opposed to other CIAO tools such as psextract) is that AE calculates the size, shape,
and position of each extraction region, and it calculates the auxiliary response file
(ARF)\footnote{http://cxc.harvard.edu/ciao/dictionary/arf.html} taking into account the PSF fraction
enclosed in the region as a function of energy.  We also use AE to refine the positions beyond the
initial {\small \sc wavdetect} estimates and calculate some useful statistics and photometry.

We briefly lay out the point source extraction process here.  First, we constructed regions to match
the PSF retreived from the CALDB library and that enclose a prescribed percentage of the PSF (90\%
default unless it needed to be adjusted to as low as 50\% for nearby sources relative to the size of
the PSF).  Then for each point source, we extracted the source events within the PSF-matched region
and a representative background in an annular region centered on the point source.

With this information, AE then provides new position estimates for each of the sources.  We refined
the positions of the sources according to the prescriptions in the AE manual.  For sources that were
$\leq 5^\prime$ from the aimpoint, we used the mean event position, and, for sources that were $>
5^\prime$ from the aimpoint, we used the correlation position.  The latter position is calculated
automatically by AE, by correlating the neighborhood around the source (not just the extracted
counts) with the source's PSF.  Since the positions of some of the sources (especially
the fainter ones) take time to converge, we ran these first few steps a minimum of 5 times.  This
provides us with very accurate source positions, useful for comparing to observations taken with
other telescopes and provides accurate flux estimates.

Lastly, we extracted the spectra for each source and its local background, which included the
creation of the ARF and RMF (redistribution matrix
file)\footnote{http://cxc.harvard.edu/ciao/dictionary/rmf.html} files for fitting the spectra.  AE
implements rules so that the background is always well-constrained.  At minimum, the background
spectrum must always have at least 100 counts and a ratio of the photometric errors of the source to
background of at least 4.0 (so that the error in the background does not dominate the
total error).  These constraints on the background extraction yielded a median background radius of
76 sky pixels with $> 99 \%$ of the sources having radii less than $\sim 200$ sky pixels and
$< 1 \%$ of the sources farthest off-axis having radii of $\sim 200$--500 sky pixels.

\subsection{Spectral Fitting and Source Properties}
\label{section:source_lists--spectral_fitting}

\begin{figure}[tbp]
\centering
\includegraphics[scale=0.36, trim = 30mm 10mm 10mm 15mm, clip]{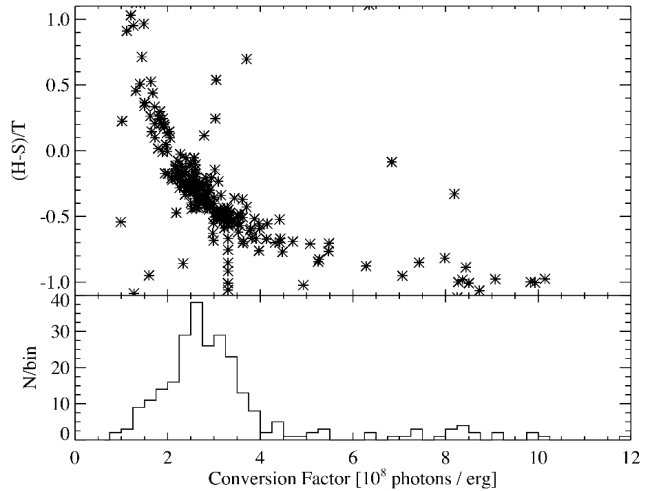}
\caption{Source hardness as a function of conversion factor from counts to ergs for sources in our
merged master source list.  The estimated source flux will be biased if a single conversion factor
is used to convert counts to energy flux units.  The vertical line of points near 3.3 represents the
spectral fits where the photon index was frozen to 1.7 and N$_\mathrm{H}$ was frozen to or floated
to the minimum value in the direction of M81.   The points only extend $\sim 1$--11 on the x-axis
because of the constraints imposed on the fit parameters, which are laid out in
\S~\ref{section:source_lists--spectral_fitting}.  The hardness ratio for a photon index of 1.7 and a
column density equal to the Galactic column density in the direction of M81 is $-0.39$ ((H$-$S)/T).
H: 2--8~keV; S: 0.5--2~keV, T: 0.5--8~keV
\label{fig:conv_factor}}
\end{figure}

\begin{figure*}[t]
\centering
\includegraphics[scale=0.35, trim = 15mm 10mm 10mm 10mm, clip]{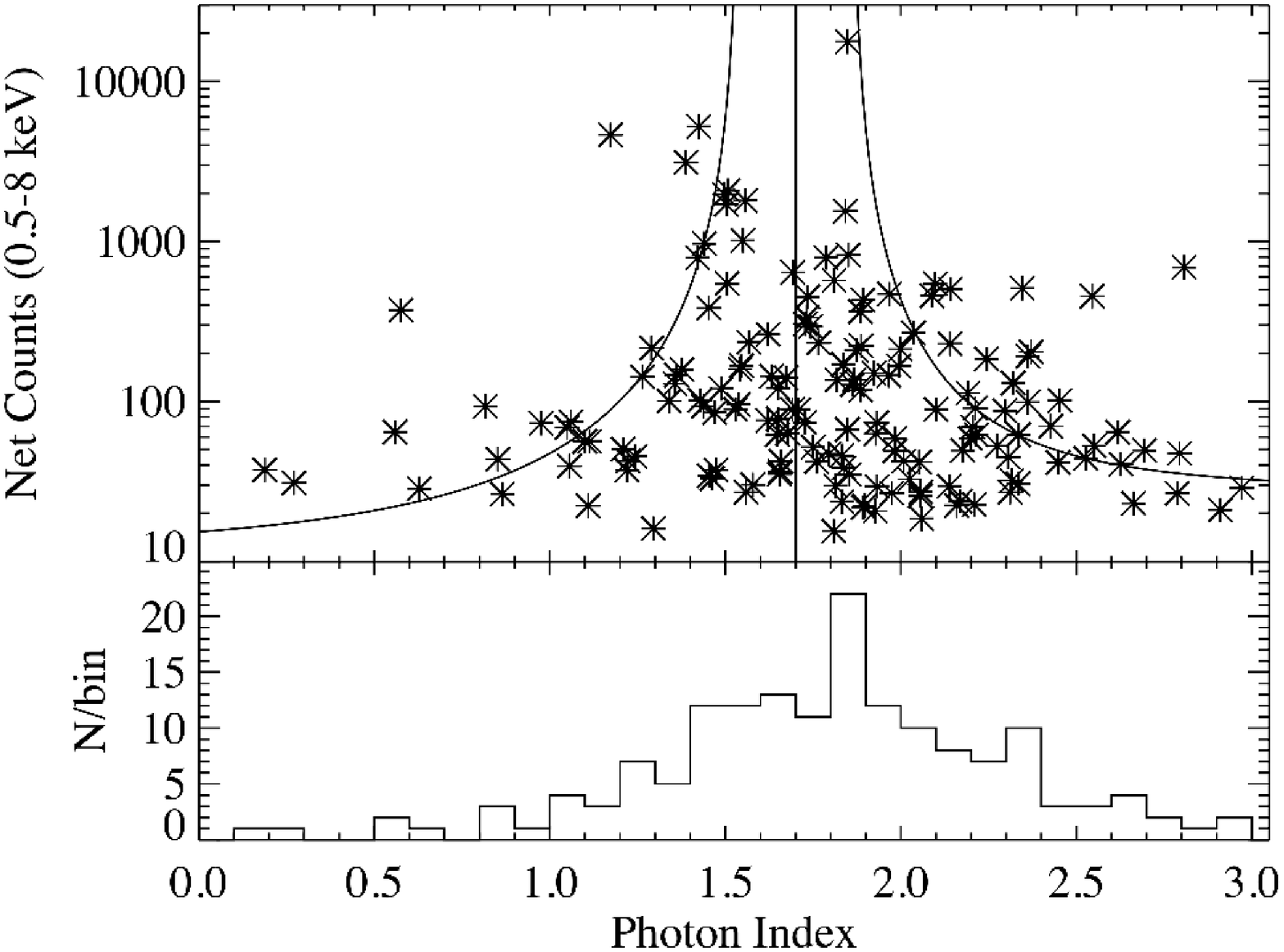}
\includegraphics[scale=0.35, trim = 15mm 10mm 10mm 10mm, clip]{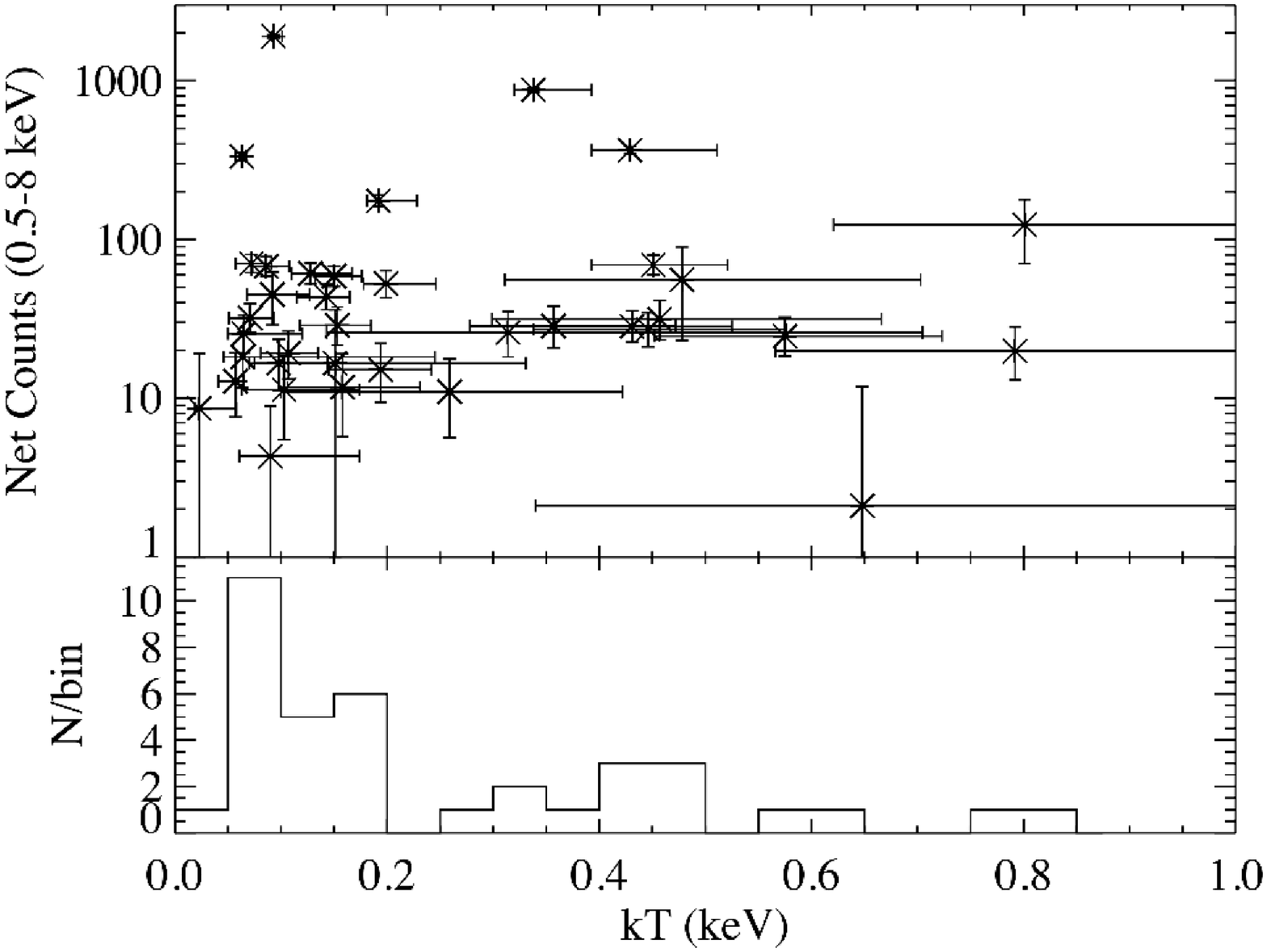}
\caption{Distributions of the two key fit parameters for the fits of the merged spectra.  The
vertical line in the upper left plot is the weighted average photon index and the two curves are
analytic fits to the $1 \sigma$ uncertainties added to a natural width of the distribution.  The
natural width, 0.32, is calculated so that 68.3\% of the points are within $1 \sigma$ of the
weighted average photon index.
\label{fig:fit_dist}}
\end{figure*}

To estimate the energy fluxes and other source properties, we fit the source and background spectrum
of each source in each observation ($\sim 4000$ spectra, most fit using automated methods detailed
below).  We tested an alternate method of estimating fluxes by using a single count-rate-to-flux
conversion factor.  We found that this would have led us to calculate different source fluxes by
factors of order unity or less (see Fig.~\ref{fig:conv_factor}; see also
\S~\ref{section:comp_M81_M31} for a brief discussion of this effect with regard to the XLFs).  We
jointly fit the unbinned source and background spectrum of each source in each
observation\footnote{For each source in each observation, the source and scaled background spectrum
fits were added together (the background was not subtracted), and then the total spectral fit was
minimized.} in Sherpa \citep{freeman01} using the C-statistic, which is similar to the \cite{cash79}
statistic but with an approximate goodness-of-fit measure, and the Powell minimization algorithm.

Since almost all of the sources have too few counts to constrain many of their source properties and
since we are mainly interested in estimating accurate fluxes, we began by fitting the source
and background spectra for each source in each observation with absorbed power-law models ({\small
\sc xswabs $\times$ xspowerlaw})\footnote{Use of the {\small \sc xsphabs} absorption model instead
does not make a significant difference to the fits.}.  We initially used the default parameter
boundaries ({\small \sc xswabs.nH}=[0.01,10] ($10^{22}$ cm$^{-2}$),
{\small \sc xspowerlaw.PhoIndx}=[-3,10], {\small \sc xspowerlaw.norm} is estimated from the data) in
Sherpa in all cases except one.  We always constrained the hydrogen column density to be at least
that of the Milky Way in the direction of M81, $4.2 \times 10^{20}$ cm$^{-2}$ \citep{dickey90}.

Since degeneracies in the fit parameters will frequently arise for very faint sources, we followed a
specialized scheme for these sources based on the number of counts in the source extraction region. 
If we extracted less than 5 counts (0.5-8 keV) for a source, the power-law index and the hydrogen
column density for the fit were frozen to 1.7 and the Galactic value, respectively.  If we extracted
more than 5 but less than 26 counts (0.5-8 keV), we froze only the power-law index to 1.7 and let
the hydrogen column density float, although, in this case, it was always poorly constrained.  For
all other sources with more than 26 counts (0.5-8 keV), we allowed all fit parameters to float.

\begin{figure*}[tbp]
\centering
\includegraphics[scale=0.8, trim = 30mm 20mm 20mm 25mm, clip]{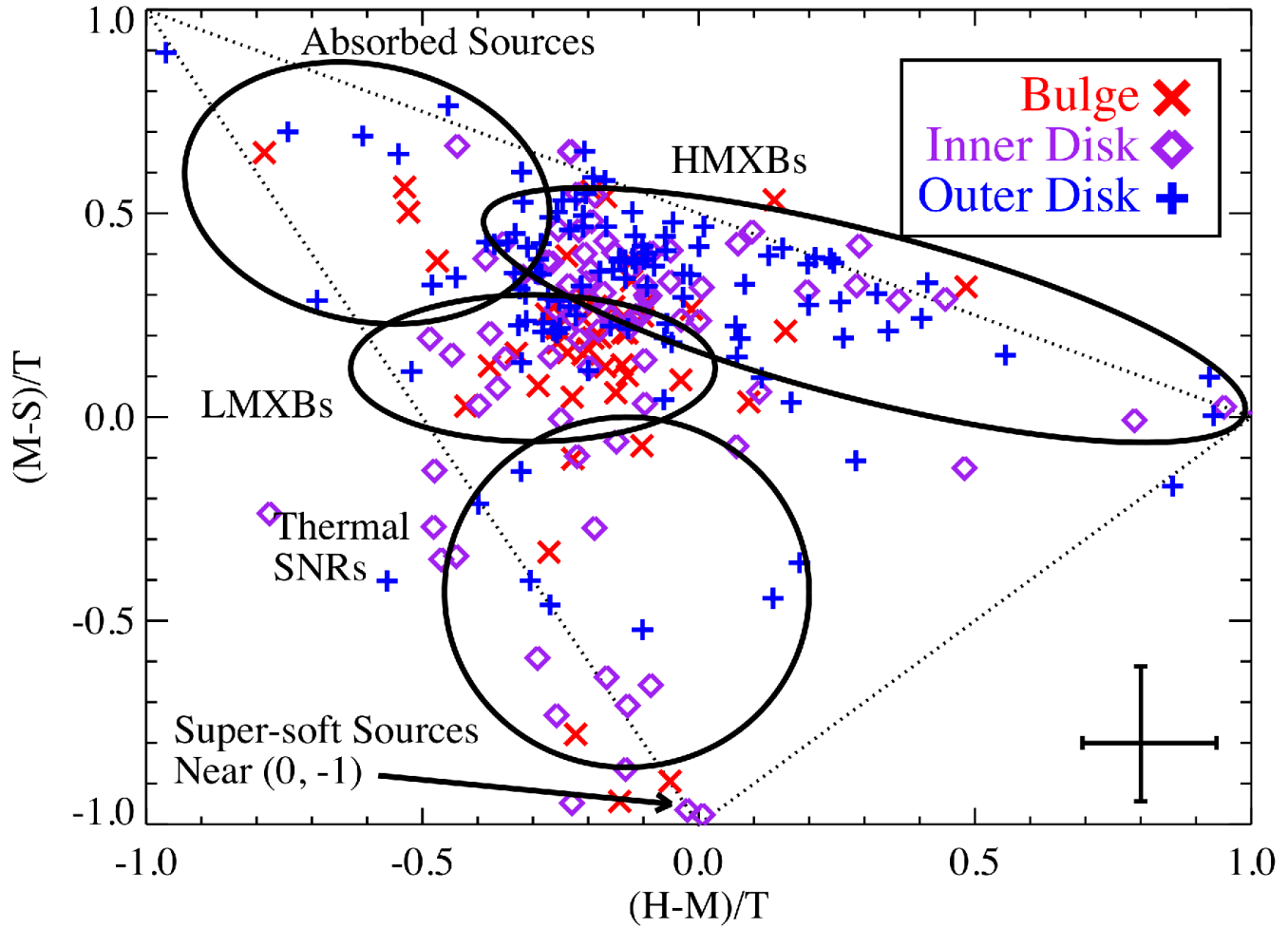}
\caption{Hardness ratios for the merged observation plotted as in \cite{prestwich03} (H: 2--8 keV;
M: 1--2 keV; S: 0.5--1 keV, T: 0.5--8 keV).  The median Gehrels error bars for all of the sources
are in the lower right corner of the plot.  The dotted lines represent the physical boundaries,
outside of which the number of net counts are negative.  This can happen for very faint sources when
the background is subtracted.  Finally, note that a large percentage of the sources in this plot are
likely background AGN (see Fig.~\ref{fig:XLFs_merged}), especially in the outer disk region, that
usually cannot be separated from the other point sources in the galaxy using this classification
scheme.
\label{fig:HR}}
\end{figure*}

Following these rules, we obtained reasonable fits for most of the sources.  However, some of the
individual-observation source spectra ($\sim 17\%$), did not have acceptable fits using these rules
alone.  First, the background spectra were not always well-fit by a single absorbed power-law model.
 The merged spectra revealed that, when well-sampled, background spectra were always quite
complicated.  These complications exhibited themselves sometimes in the individual observation
source spectra.  For the cases where the reduced background goodness of fit statistic was
$\ge 0.9$\footnote{The cutoff value for the goodness of fit statistic is smaller than 1 because
this reduced statistic was typically underestimated near $\sim 0.5$.}, a more complicated
background model, two power laws and a blackbody ({\small \sc xswabs $\times$ (xspowerlaw $+$
xspowerlaw $+$ xsbbody)}) was used to achieve a good fit.  This ad hoc combination of models is only
used to model the shape of the background spectra.

Second, there were some sources ($\lesssim 4$\% of the fits) in which the best-fit hydrogen column
density of the source was found at the maximum of the parameter space, $10^{23}$ cm$^{-2}$.  Since
the simple absorption model does not account for multiple photon scatterings through a Compton-thick
medium and we did not consider other absorption models for our spectral fitting, we refit the
spectra using a maximum hydrogen column density of $10^{24}$ cm$^{-2}$, which allowed for
reasonable spectral fits for most of these sources.  However, there were still a small subset of
sources ($\lesssim 1$\% of the fits) with 5--25~counts where the best-fit hydrogen column densities
were at $10^{24}$ cm$^{-2}$.  For these few sources, we allowed the power-law photon index,
$\Gamma$, to float.  This led to resonable fits for these sources.

Third, there were sources with best-fit power-law indices up to 10, the maximum parameter space
boundary.  For these soft sources with $\Gamma > 3$ ($\lesssim 3$\%), we changed the source model to
a simple thermal ({\small \sc xsbbody}) model and achieved good fits (C-statistic $\le 1$ with kT
$<$ 1 keV).  Our change in spectral model here does not imply that there are not sources that could
be well-fit by this model with temperatures in excess of 1~keV.  Given the limited number of source
counts for most sources, we expect that such sources were well-fit by the absorbed power-law model.
This only implies that there were sources that could not be well-fit by the absorbed power-law model
within a reasonable range of photon indices because their spectra were too soft.  This result
indicates that we cannot differentiate these models for most of our sources that are not very soft.
This is acceptable since our goal is not to differentiate source models.  These comments also apply
to the merged spectra that are mentioned below.

Fourth, a very small number of sources ($\lesssim 1$\%) had $\Gamma < 0$ or had spectra such that
Sherpa could not find a reasonable local minimum and frequently ran into parameter space boundaries.
These were generally sources that had only a few counts, and were located in regions where the
background spectrum was complicated.  In these cases, the software had to be manually guided until a
reasonable fit was found.

In summary, we allowed for fairly liberal upper and lower bounds for the fit parameters because
there are a wide range of different types of sources in our sample (e.g., LMXBs, HMXBs, SNRs, SSSs,
ULXs, background AGN, etc.).  After considerable experimentation and iteration, the following
constraints were imposed on our spectral fitting process:

\begin{enumerate}

\item We required the reduced goodness of fit statistic for the source to be better than 1.2.

\item We required the reduced goodness of fit statistic for the background to be better than 0.9 for
sources within 7~arcmin of the aimpoint and better than 1.4 for sources farther off-axis than
7~arcmin.

\item The source absorption, N$_\text{H}$, must be less than $10^{24}$~cm$^{-2}$ (the Compton-thick
limit).

\item No sources can be very near to or stuck at parameter space boundaries.

\item The source power-law fits cannot have a photon index greater than 3 or less than 0.

\item The blackbody (thermal) fits were specifically implemented only for very soft sources and do
not have kT greater than 1~keV.

\end{enumerate}
At the end of the fitting process, all fits abided by these rules.

Out of all source fits in each of the 16 observations, two sources in ObsID 735 required special
attention with the use of more complicated models.  First, the brightest ULX (source 21) was the
only source that suffered from significant pileup and, therefore, Sherpa's JD pileup model
\citep{davis01} was used in addition to the power-law fit.  The second source, SN1993J (the aim of
ObsID 735), was fit by two low-temperature absorbed thermal emission-line (vmekal) components.
Since care has already been taken in fitting these sources \citep{swartz03}, we used these models.

We also fit the source spectra of ObsIDs 5935--5949, the merged observation, to better understand
the properties of each source.  We did not include the ObsID 735 in the merged spectra because the
ARF changes significantly and there appears to be spectral variability in at least a few sources
(see \S~\ref{section:spectral_var}).  We fit the merged source spectra with simple absorbed
power-law models as above.  We found that 219 of the 262 sources were well-fit by this method
(CSTAT~$\lesssim$~1).  Again, two sources (the brightest ULX and SN1993J) were fit with special
models, as described in the previous paragraph.  The remaining sources were better fit by simple
blackbody (thermal) models.  Almost all of the sources better fit with the simple thermal model also
had hardness ratios indicative of thermal SNR or SSS populations (see \S~\ref{section:HR}).   In all
cases, the background was poorly fit by a single absorbed power law.  We used the more complicated
background model expressed above, which achieved good fits.

Fit parameter distributions are shown in Figure~\ref{fig:fit_dist}.  There appears to be no
correlation between the net counts and the blackbody temperature for the softest sources, although
there is a lower cutoff in the blackbody temperature, which is likely due to the dropoff in
sensitivity of Chandra at lower energies and foreground absorption.  The distribution of photon
indices can almost entirely be accounted for by the uncertainties in the photon indices
($\sim 68.3\%$ within $1 \sigma$ of the mean) with a natural width of 0.32 to the distribution.

For each source in the master and borderline source lists, we compile one table for our merged
source photometry and another for our merged spectral fits, which are both ancillary tables.

\section{Hardness Ratios}
\label{section:HR}

Following \cite{prestwich03}, we calculate hardness ratios using the AE pipeline for the sources in
our master source list (Fig.~\ref{fig:HR}) from the background-subtracted counts of the merged
observation in different bands (H: 2--8 keV; M: 1--2 keV; S: 0.5--1 keV, T: 0.5--8 keV).  Since most
of the sources have far too few counts to make significant statements about the source properties,
we use these hardness ratios to estimate the spectral properties of the differing source populations
in this dataset.  The population of sources shows the full range of expected colors as seen in
Fig.~\ref{fig:HR}.

This classification is based on the general characteristics of the HMXB and LMXB populations in our
Galaxy.  The former are predominantly pulsar X-ray binaries and hence have hard spectra, while the
latter host either black holes or low-magnetic field neutron stars resulting in softer spectra (at
least at the luminosities that we are probing with these observations).  HMXBs with black-holes
could also be in the same locus but, in our Galaxy, these are substantially rarer than LMXBs.  Given
the similarity of M81 with our Galaxy, we would also expect that the vast majority of the objects in
this locus are LMXBs.  As was emphasized in \cite{prestwich03}, the color of an individual source
calculated in this fashion is not sufficient to determine the precise nature of the source.
Instead, this color classification scheme is useful primarily for population studies, and we take
advantage of this fact in the next section.

Further insight into the multi-wavelength properties of these sources requires detailed source
matching with additional observations (e.g., optical observations), which is beyond the scope of
this study.  As a result, we cannot individually identify background AGN apart from galactic source
populations in M81 at the present time.  To varying degrees, this sometimes limits the
interpretation of our:  color-color analysis in this section, variability analysis in
\S~\ref{section:indiv_src_var}, and interpretation and comparisons of the disk XLFs (mainly for the
outer disk; \S~\ref{section:XLFs}).  Also, because of this and the fact that our monitoring
observations mainly focus on the bulge and inner disk regions, we do not perform a spatial
correlation analysis between the positions of the sources in M81 to the spiral arms of M81, which
has been previously done \citep{swartz03}.  Some of these tasks will be carried out in a future
publication that will include detailed source matching with the deepest
\emph{Hubble Space Telescope} image to date (Zezas et al. in prep.).  Despite these current
limitations, we do carry out interesting spatial, variability, and XLF analysis in the later
sections.

\subsection{Separating the Bulge and Disk}
\label{section:HR--bulge_disk_sep}

\begin{figure}[tbp]
\centering
\includegraphics[scale=0.39, trim = 30mm 20mm 20mm 25mm, clip]{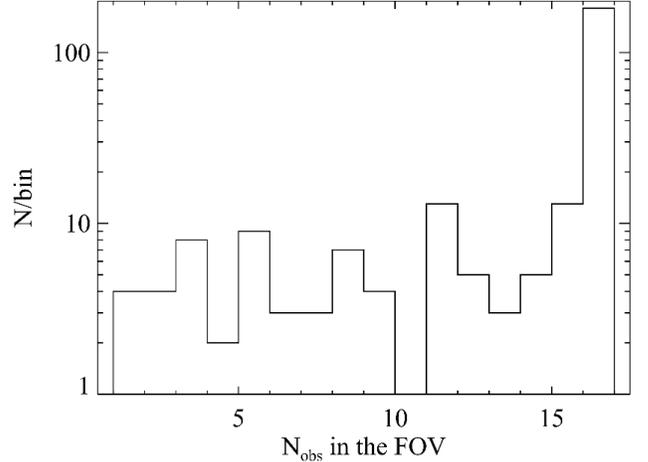}
\caption{The number of observations that each source in the master source list in the field of view.
There are a number of situations throughout this work where excluding sources not in the field of
view of all 16 observations is appropriate for consistency.
\label{fig:src_coverage}}
\end{figure}

In order to search for trends in the nature of X-ray point sources in nearby galaxies, it is common
practice to divide them into two primary groups, bulge sources and disk sources.  However, in any
disk galaxy, such a clear division almost never exists because the stellar populations rarely are
isolated from one another \citep[e.g.,][]{kormendy10}.  M81 is an early-type, spiral galaxy with a
considerable, predominantly old stellar population in the bulge that blends into the inner parts of
the disk.  The inclination of M81 \citep[$\sim 35^{\circ}$;][]{boggess59} exacerbates the blended
appearance.  Nevertheless, previous works have attempted to define a single separation between the
bulge and the disk of this galaxy.

For example, in the analysis of ObsID 735, \cite{tennant01} defined the separation of the bulge and
the disk roughly based on the morphology of the galaxy (a
$4.7 \times 2.35$\footnote{All ellipse lengths represent the diameters along the major and minor
axes of the ellipses.} arcmin ellipse with a major axis position angle of 149 degrees).  Later, in
the more detailed study of this dataset, \cite{swartz03} chose a larger, physically-motivated
separation between the bulge and disk of the galaxy using the inner Lindblad Resonance (a
$7.64 \times 3.94$ arcmin ellipse with the same position angle).  However, these are just two
examples of how one can separate these regions, and there are many other ways that this can be done.
For instance, $R$-band or $H\alpha$ isophotal fits or $U-B$ color changes all yield different
results (Jay Gallagher---private communication).

We propose to separate the bulge and disk in a different way than listed above by taking advantage
of the classification scheme laid out by \cite{prestwich03}.  We can use this diagram as a guide to
separate the bulge and disk of this galaxy since certain populations of sources tend to be more
strongly-associated with different parts of the galaxy.  For example, although we expect to find
LMXBs in all regions of the galaxy, we expect a large fraction of the sources in the bulge region to
be LMXBs since older stellar populations dominate here.  Also, we expect to see very few or no HMXBs
or SNRs in the bulge region since these sources should only be found in regions of active star
formation, primarily the disk.

By taking different inclination-corrected radial cuts, we can find at which radius sources with
colors consistent with LMXBs and HMXBs dominate or when they are hardly present at all.  Following
this technique, we define the ``bulge" to include all sources inside a $4 \times 2$ arcmin ellipse
at a position angle of 149 degrees with respect to the major axis, which is slightly smaller than
the morphology-based definition in \cite{tennant01}.  We define the ``outer disk" to be all sources
outside a $11 \times 5.5$ arcmin ellipse with the same position angle, but within the hatched
regions of Fig.~\ref{fig:outline} ($\sim 41$~arcmin$^2$), which is closer to but considerably larger
than the disk as defined in \cite{swartz03}, based on the inner Lindblad Resonsance.

This method leaves an undetermined inclination-corrected annular region of the galaxy, which we
refer to as the ``inner disk" region.  This region includes sources from all sections of the
color-color plot and is consistent with the properties of both the bulge and the disk of the galaxy.
 The apparent properties of the incompleteness-corrected XLFs are also consistent with the
properties of both the bulge and the disk of the galaxy, although the fits suggest that the XLF is
very disk-like (\S~\ref{section:XLFs}).

\section{Individual Source Variability}
\label{section:indiv_src_var}

\begin{figure*}[tbp]
\centering
\includegraphics[scale=0.38, trim = 35mm 20mm 10mm 10mm, clip]{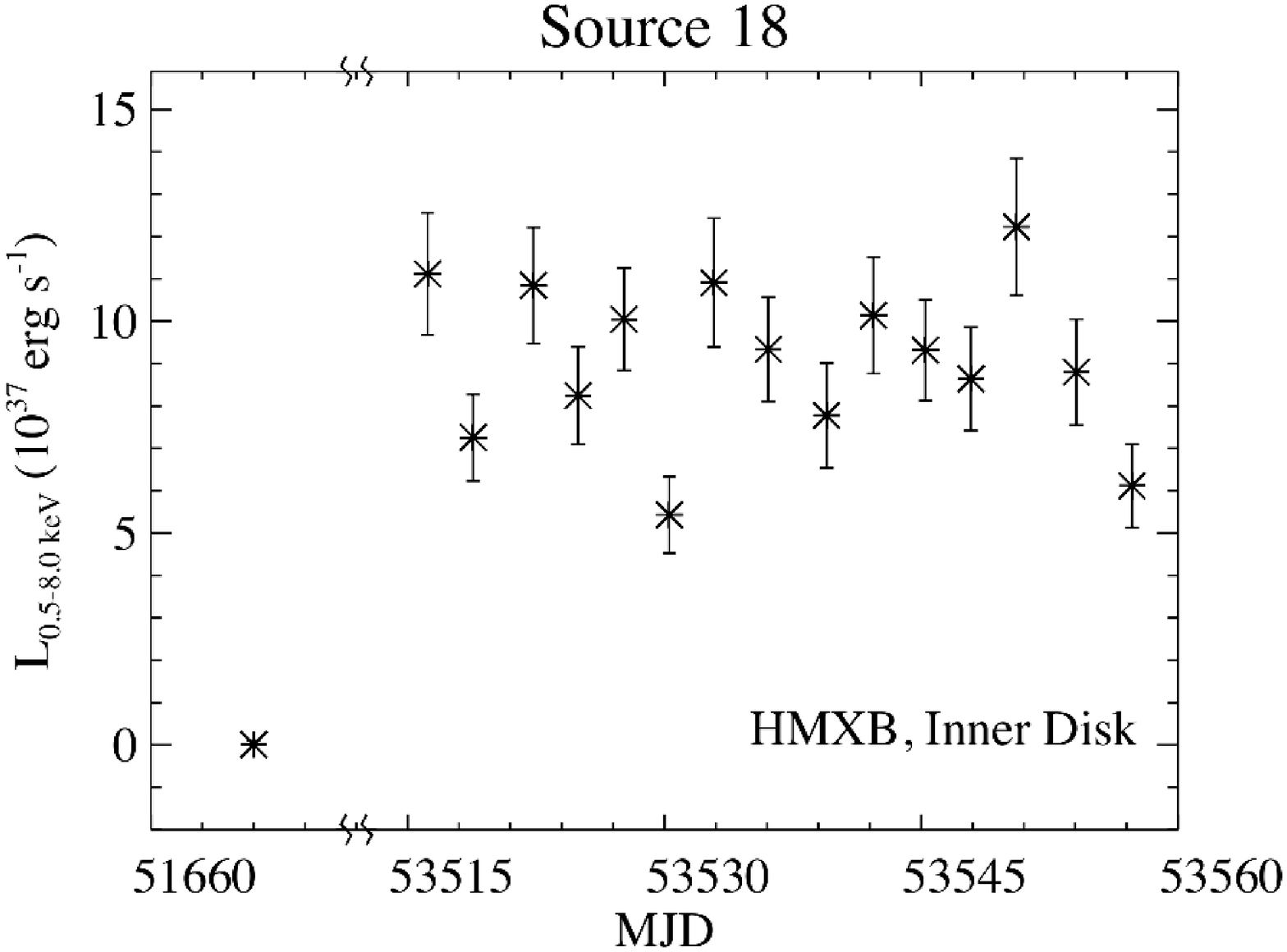}
\includegraphics[scale=0.38, trim = 35mm 20mm 10mm 10mm, clip]{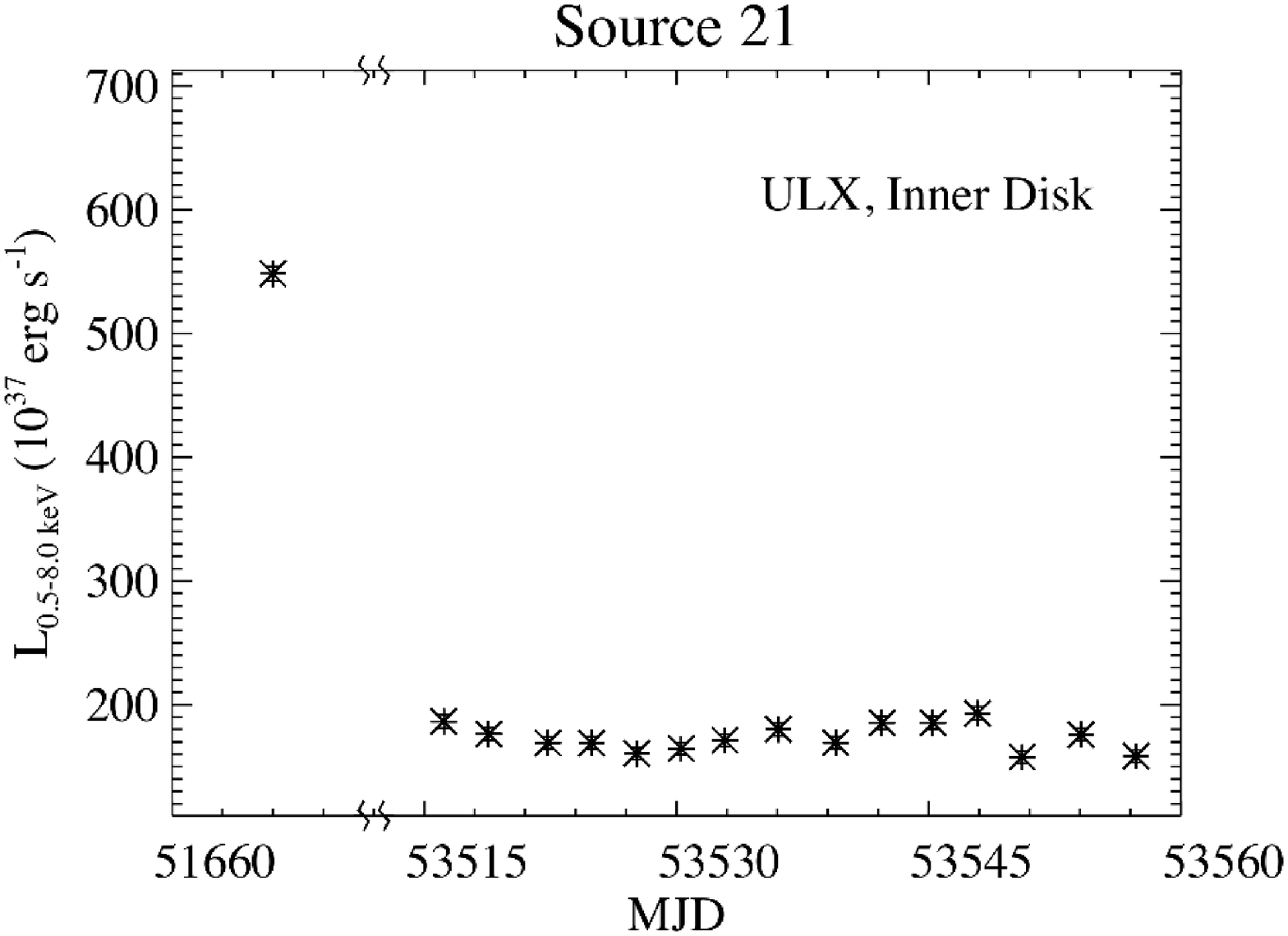}
\includegraphics[scale=0.38, trim = 35mm 20mm 10mm 10mm, clip]{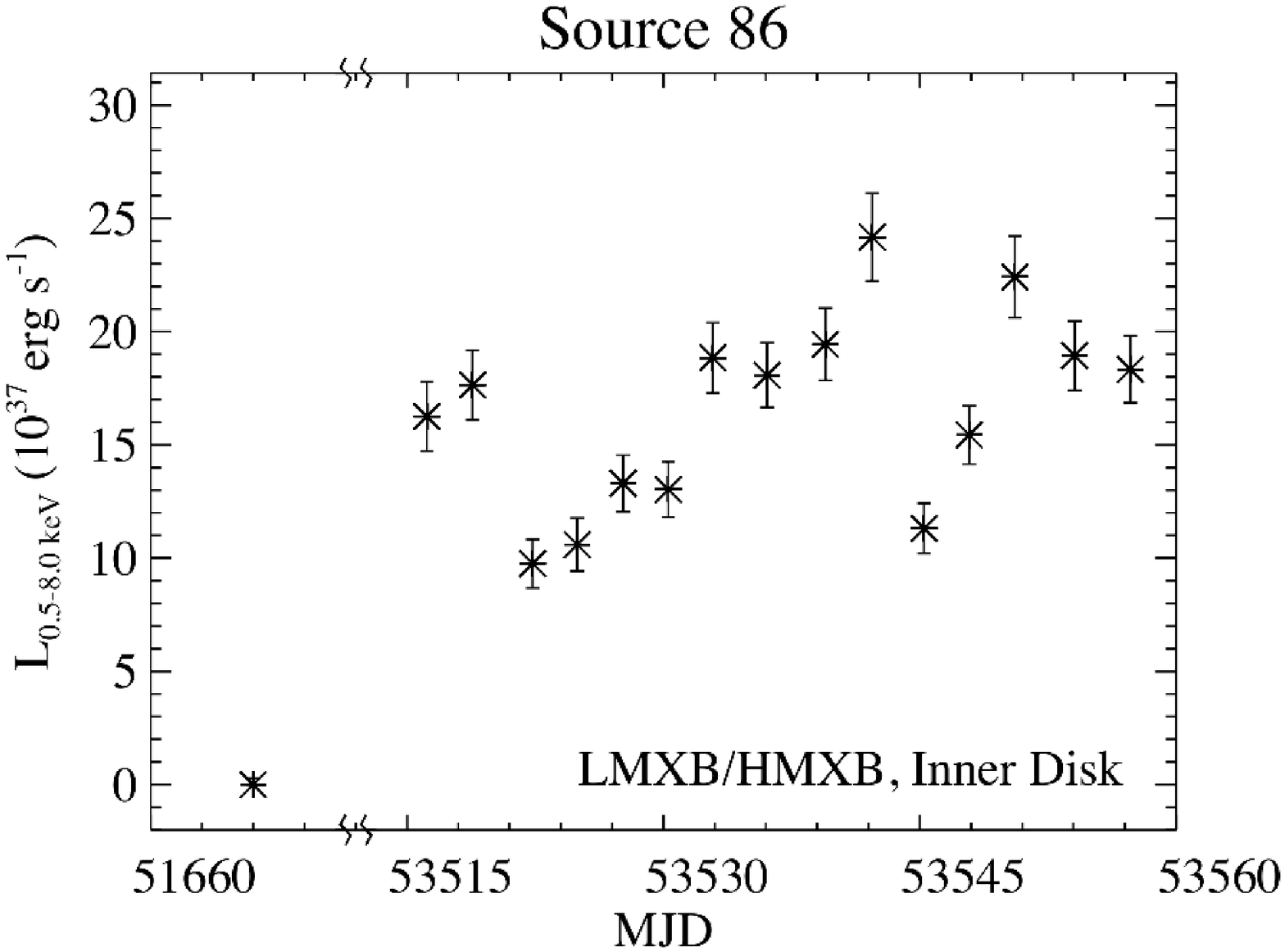}
\includegraphics[scale=0.38, trim = 35mm 20mm 10mm 10mm, clip]{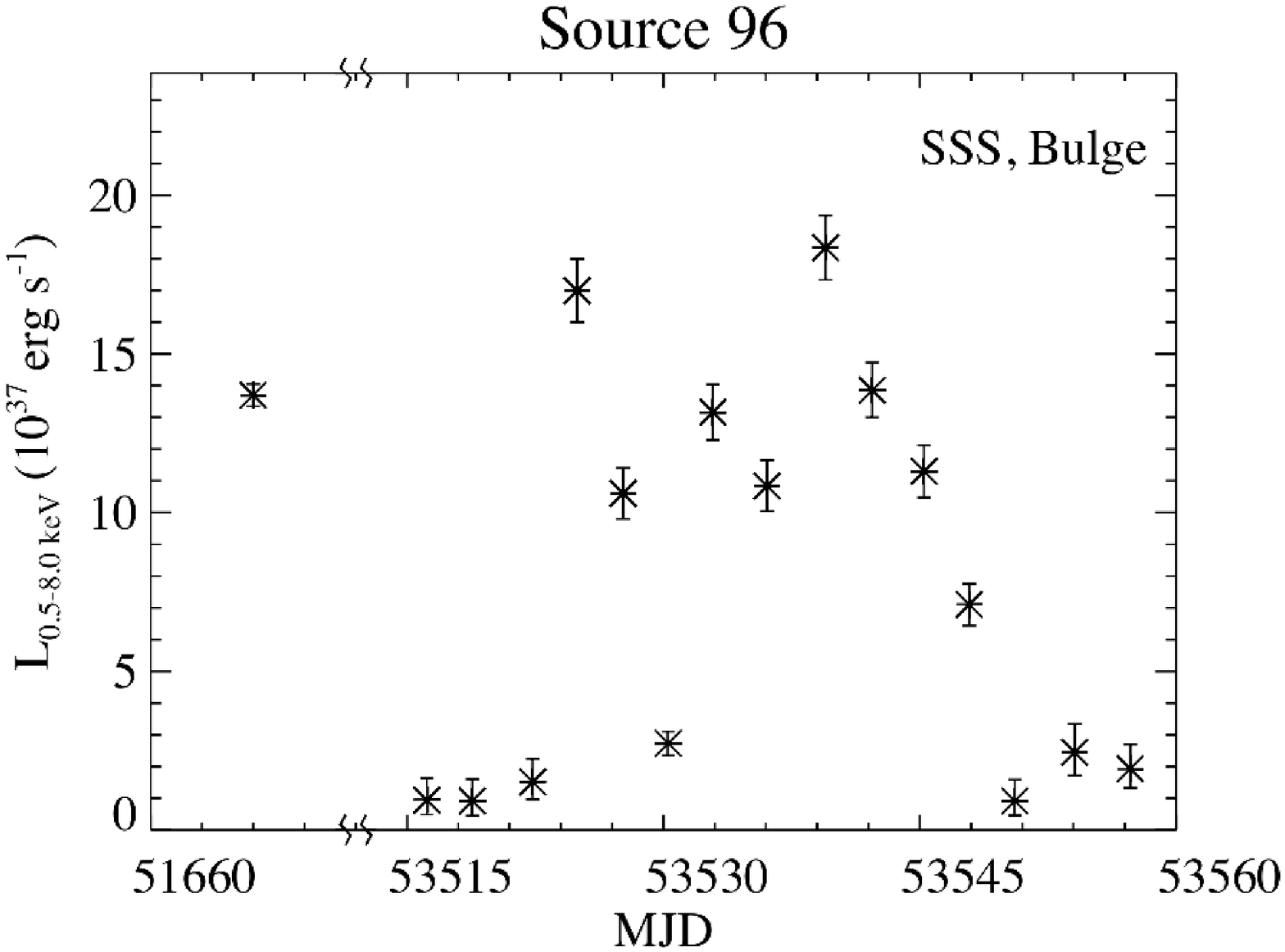}
\includegraphics[scale=0.38, trim = 35mm 20mm 10mm 10mm, clip]{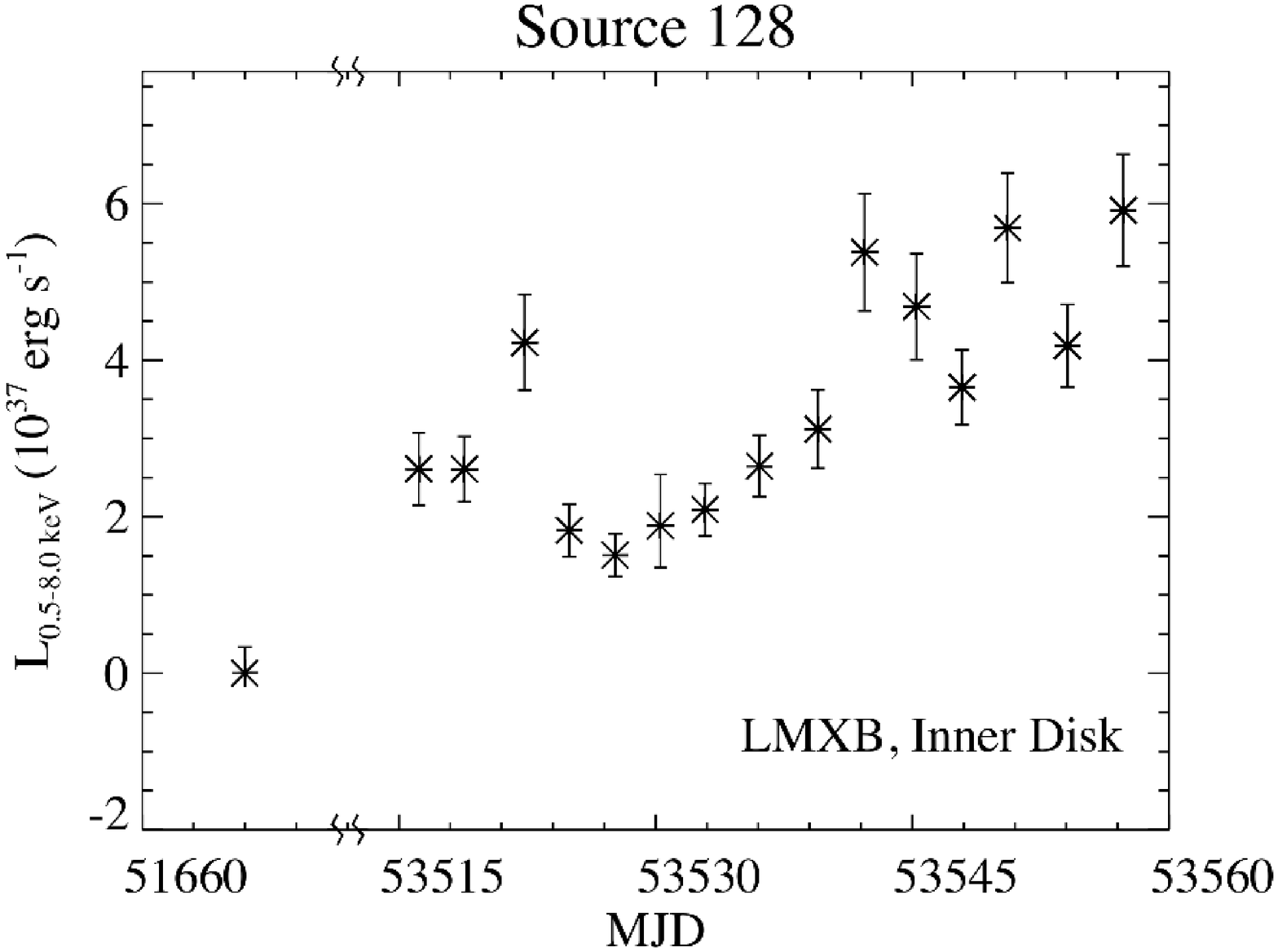}
\includegraphics[scale=0.38, trim = 35mm 20mm 10mm 10mm, clip]{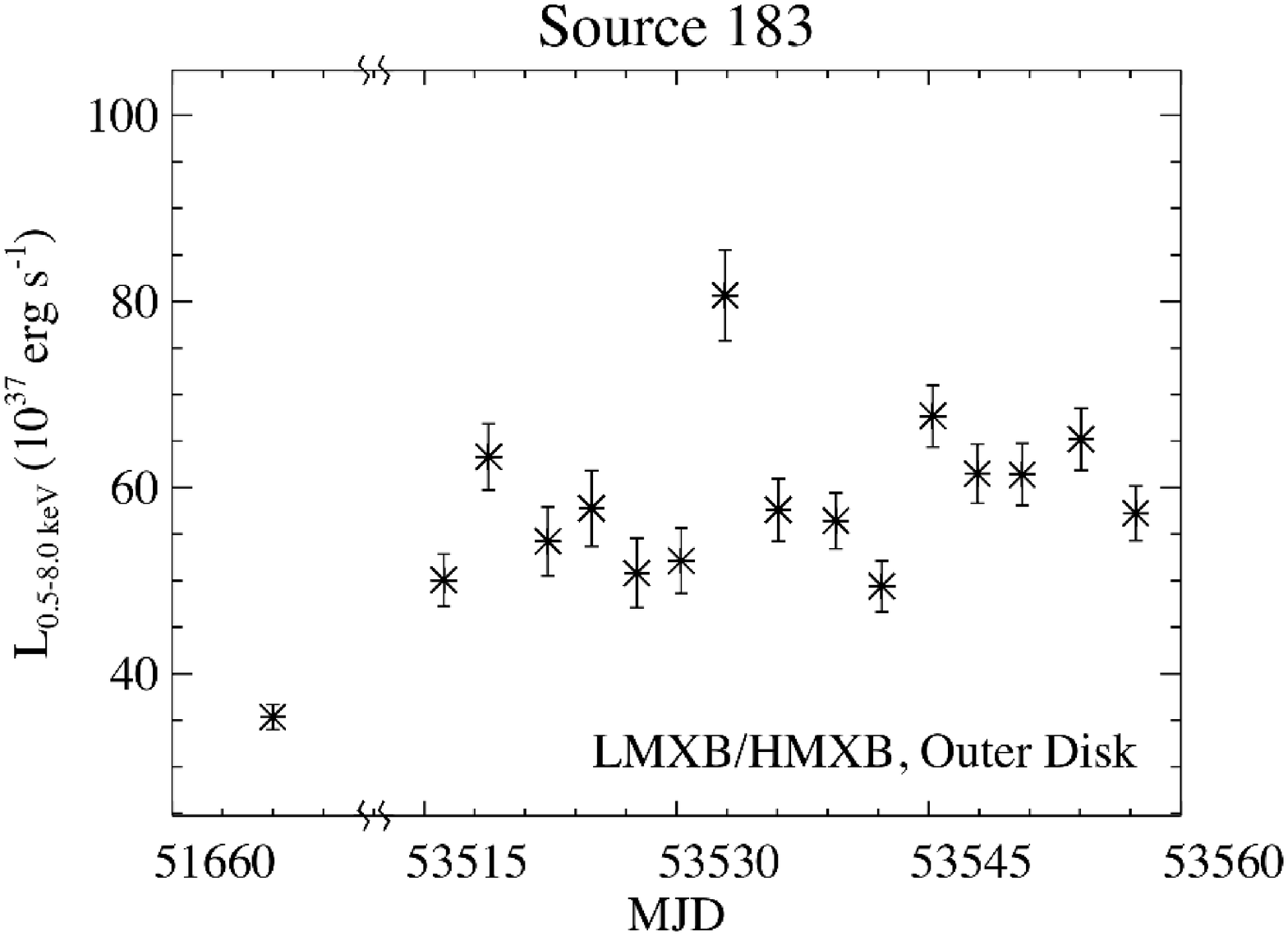}
\caption{Six sources with the most-significant variability.  The most-significant variability, as
parametrized by equation~\eqref{eq:var}, always occurs on the 5-year timescale in these cases
(comparing the weighted average luminosity and uncertainty of the merged observation, to the
luminosity and uncertainty of ObsID 735) . Note the breaks in the time axes.  The luminosities were
calculated from each of the spectral fits.  The errors in the luminosity were estimated from the
Bayesian error in the counts \citep{kraft91} and then scaled appropriately.  We include labels that
indicate the region of the galaxy in which each source is located and the ``best guess" of the type
each source from Fig.~\ref{fig:HR} (the color of an individual source is not sufficient to determine
the precise nature of the source).
\label{fig:indiv_var}}
\end{figure*}

\subsection{Flux Variability}
\label{section:flux_var}

Since one of the primary goals of this study is to test the significance of source variability on
the XLFs, we need to first estimate the level of variability that we can detect in the individual
source population.  We parameterize individual source variability by comparing the difference in
luminosity for each source in each observation with its corresponding uncertainties.  Since we probe
different timescales, we parameterize the variability in two ways.

First, for each source, we calculate the significance statistic between each of the individual
ObsIDs 5935--5949, which probes the days--weeks timescales.  We use the same variability
parameterization as in \cite{fridriksson08}:

\begin{equation}
\label{eq:var}
S_\mathrm{flux} = \text{max}_{i,j} \frac {|F_i - F_j|} { \sqrt{ \sigma^2_{F_i} + \sigma^2_{F_j} } },
\end{equation}
where the fluxes ($F_i,F_j$) were calculated as described in
\S~\ref{section:source_lists--spectral_fitting}.  We estimate the uncertainty in the flux
($\sigma_{F_i}, \sigma_{F_j}$) from the 90\% confidence interval in the counts \citep{kraft91}, and
then take the average of the uneven Poisson uncertainties to form a single uncertainty as required
by equation~\eqref{eq:var}.  Second, we calculate the significance statistic on the 5-year timescale
by comparing the weighted average fluxes and appropriately propagated uncertainties of the merged
observation to the fluxes in ObsID~735 using equation~\eqref{eq:var}.

We consider sources with $S_\mathrm{flux} > 3$ to have significant variability.  For sources that
are in the field of view of all observations (see Fig.~\ref{fig:src_coverage}), we find that 
16\% of them exhibit significant variability on the days--weeks timescales and 
25\% of the sources exhibit significant variability on the 5-year timescale.  For some sources, we
find considerable variability as much as approximately an order of magnitude in luminosity.
However, we likely have missed substantial variations in some sources, especially some of the
fainter ones, because of limited signal-to-noise and how we search for variability.  Thus, the
fraction of the population that we measure as variable is a lower limit.

We plot light curves of the 6 sources with the strongest variability in either of the two different
timescales ($S_\mathrm{flux} > 11$; Fig.~\ref{fig:indiv_var}).  In these 6 most extreme cases, the
most significant variability always occurs on the 5-year timescale (at least in part due to the
smaller errors in the luminosities associated with this timescale comparison).  In addition, all of
these sources have colors in Fig.~\ref{fig:HR} consistent with LMXBs and HMXBs ((H-M)/T$ =
-0.5$--0.0 and (M-S)/T = 0.0--0.5) except for source number 96, which is consistent with a SSS.

For the remainder of the variable sources, we see a wide range in the level and timescale of
variability.  However, there appear to be groups of sources with similar variability
characteristics.  For instance, there is a group of sources ($\sim 20$) that have luminosities on
the order of a few times 10$^{36}$~erg~s$^{-1}$ or less for most of the observations, but that
brighten by about an order of magnitude over 1-3 observations or a timescale of a few days.  In
particular, six of these sources, which all have colors consistent with HMXBs, brighten by a factor
of 5--30 over one of ObsIDs~5935-5949 ($\sim 2.5 \sigma$).  One of these sources is in the outer
disk region and five are in the inner disk region.  Sources undergoing an outburst like these could
be population of massive star transient sources (e.g., Be star binaries) like those in the
Magellanic Clouds and our Galaxy \citep[e.g.,][]{liu06,meurs89}.

\subsection{Spectral Variability}
\label{section:spectral_var}

We also tested for spectral variability by constructing the same variability statistic as in
equation~\eqref{eq:var} for the column density and power-law index or blackbody temperature.  The
fluxes in this equation are replaced by the best-fit values of these parameters and the flux
uncertainties are replaced by the uncertainties from the spectral fits.

In the master source list, significant spectral variability based on the column density was found
for only one source, number 21 (the brightest source and ULX).  The variability was found only for
the comparison of the merged observation and ObsID 735 for the 5-year timescale.  Significant
spectral variability based on the power-law index was found for only four sources:  5, 6, 19, 22.
The most significant variability for sources 5 and 22 occurs on the days--weeks timescale and, for
sources 6 and 19, occurs on the 5-year timescale between single observations.  There were no sources
with significant variability in the blackbody temperature.

All of these sources have colors consistent with LMXBs and HMXBs ((H-M)/T$ = -0.4$--0.0 and (M-S)/T
= 0.1--0.6).  Detecting significant spectral variability in only a few cases is not unexpected given
the very low signal-to-noise ratio for most sources in the individual observations.  Finally, as was
the case for the flux variability, we note that the amount of spectral variability only represents a
lower limit to the spectral variability.

\section{Luminosity Functions}
\label{section:XLFs}

\begin{deluxetable*}{lcccc}
\tablecaption{Lower Luminosity Cutoffs for the XLF Comparisons}
\tablehead{
	\colhead{Comparison} &
	\multicolumn{4}{c}{Luminosity Cutoff ($10^{37}$~erg~s$^{-1}$)} \\
	\colhead{Type} &
	\colhead{Entire Galaxy} &
	\colhead{Bulge} &
	\colhead{Inner Disk} &
	\colhead{Outer Disk} }
\startdata
ObsIDs 5935--5949 (days--weeks) & 2.9 & 1.3 & 2.6 & 2.9 \\
ObsIDs 5935--5949 to 735 (5-year) & 5.1 & 1.7 & 5.1 & 3.8 \\
Merged to ObsID 735 (5-year) & 1.8 & 0.33 & 1.5 & 1.3 \\
ObsIDs 5935--5949 to Merged & 1.5 & 0.80 & 1.5 & 1.5 \\
\enddata
\tablecomments{We include the comparison of the merged to ObsID 735 XLF because it goes considerably
deeper.}
\label{table:XLF_cutoffs}
\end{deluxetable*}

\begin{figure*}[t]
\centering
\includegraphics[scale=0.35, trim = 30mm 10mm 5mm 20mm, clip]{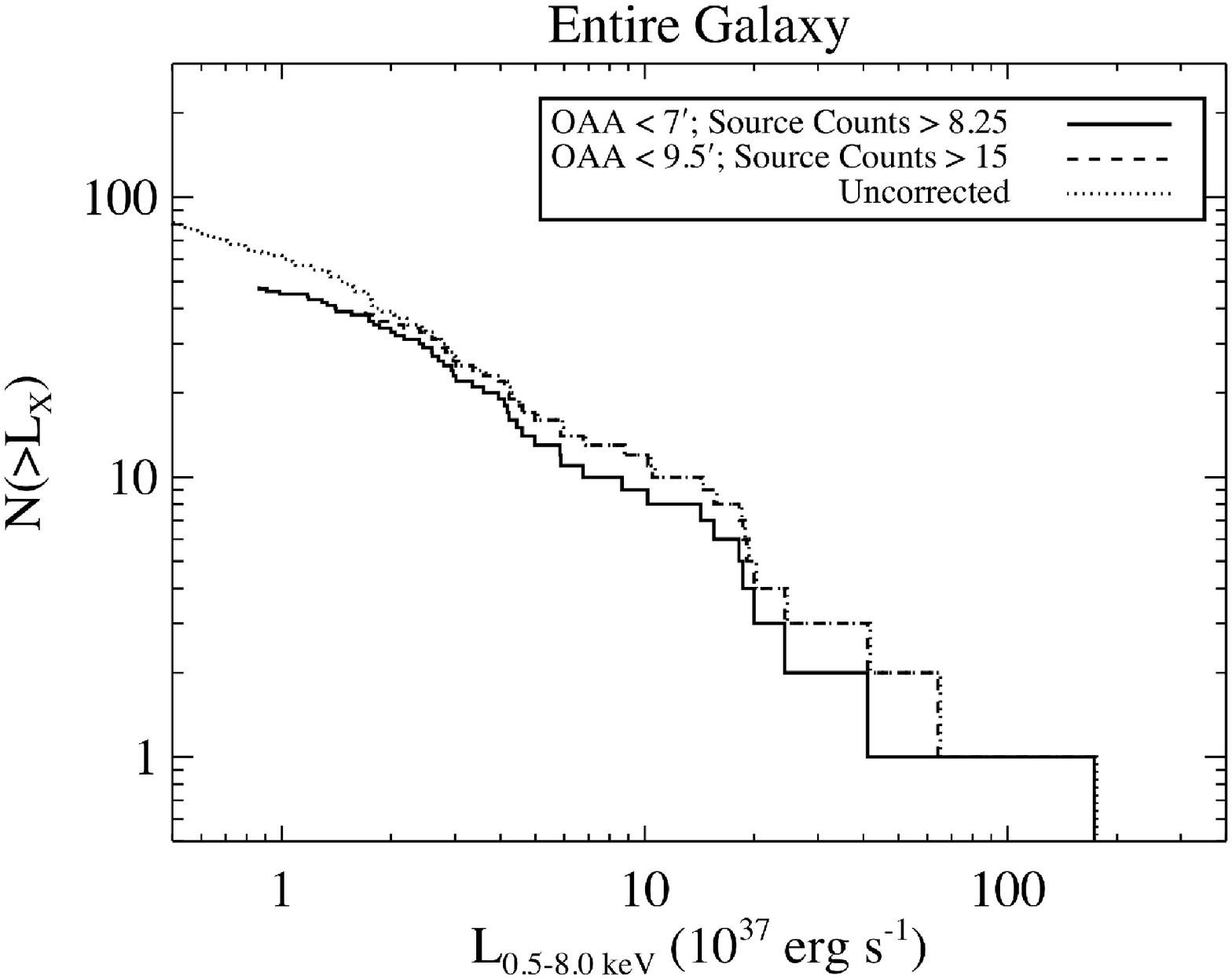}
\includegraphics[scale=0.35, trim = 30mm 10mm 5mm 20mm, clip]{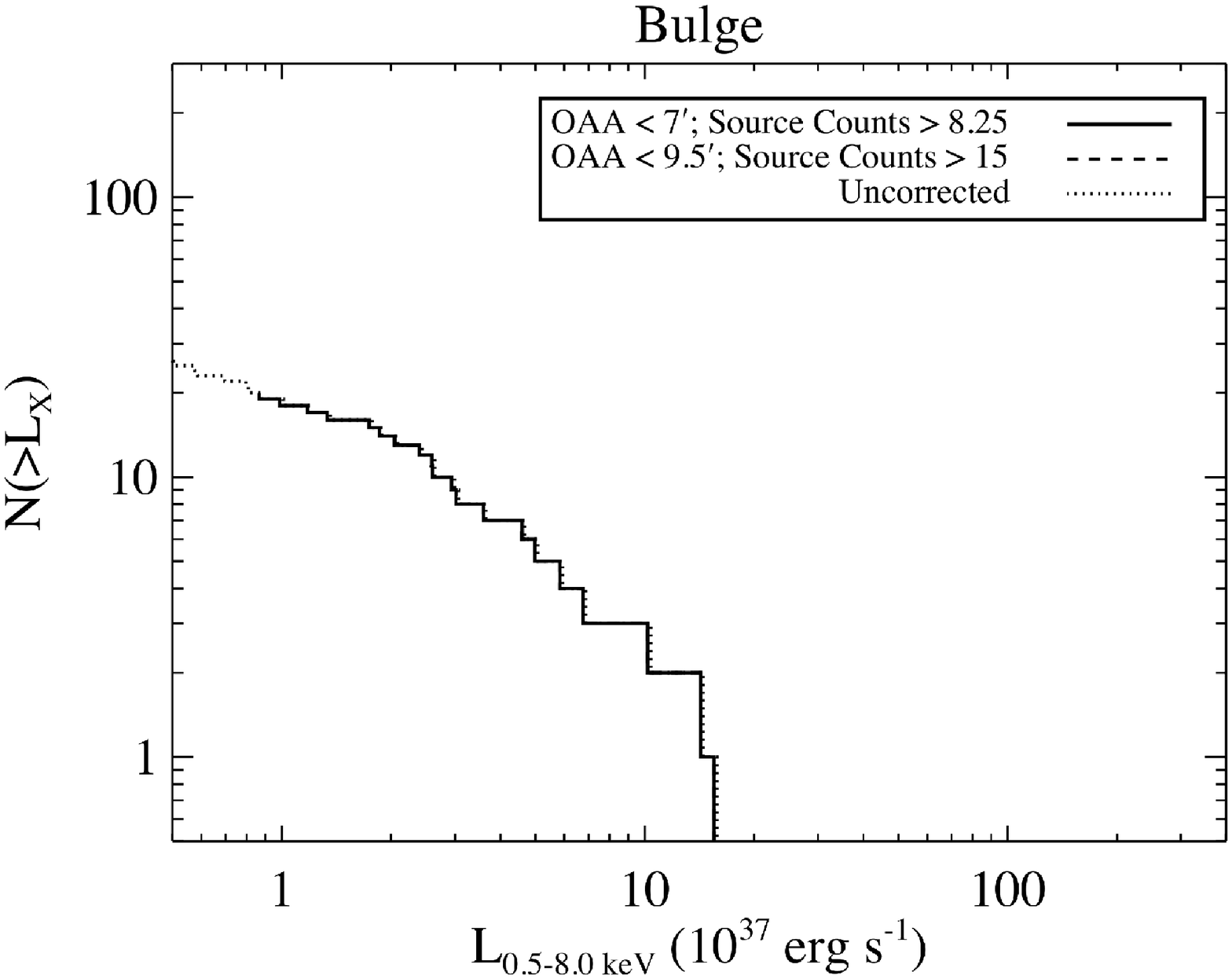}
\includegraphics[scale=0.35, trim = 30mm 10mm 5mm 20mm, clip]{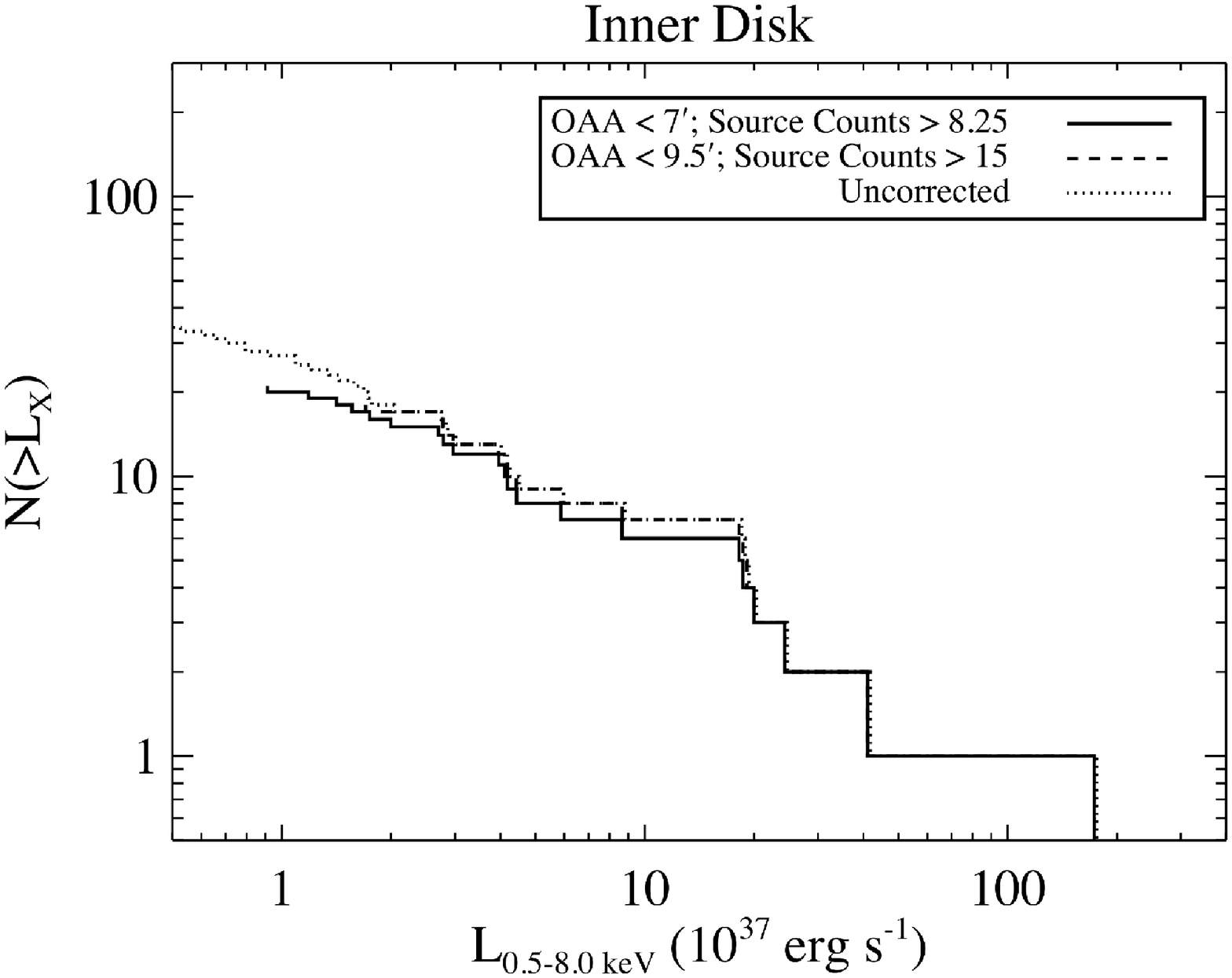}
\includegraphics[scale=0.35, trim = 30mm 10mm 5mm 20mm, clip]{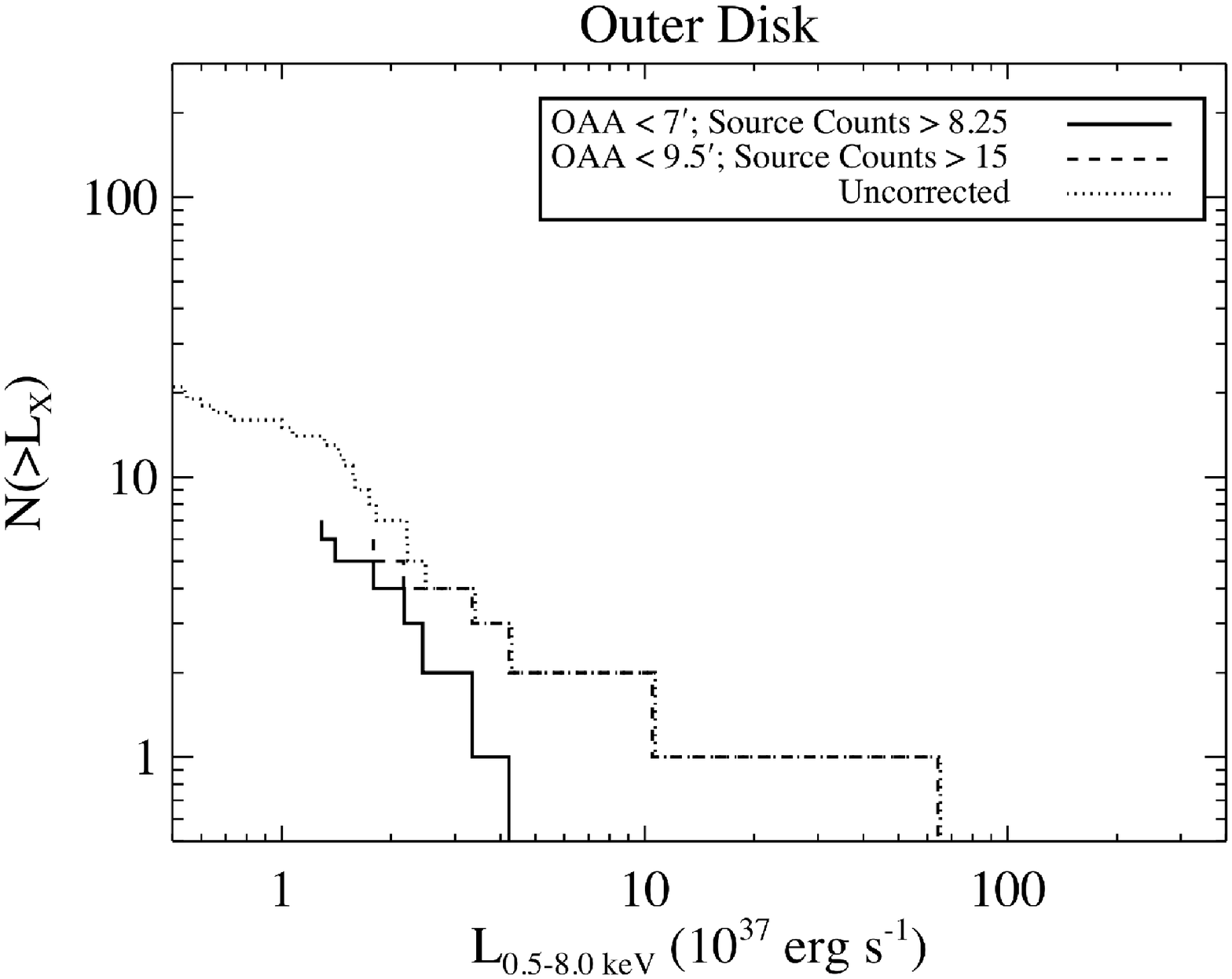}
\caption{Example cumulative XLFs for ObsID 5948 for sources that are in the field of view of all 16
observations.  Each plot represents a different region of M81.
\label{fig:XLFs_5940}}
\end{figure*}

We construct XLFs of each region of the galaxy (all regions, bulge, inner disk, and outer disk;
see \S~\ref{section:HR}) for each observation (ObsIDs:  5935--5949, 735; Fig.~\ref{fig:XLFs_5940})
and for the merged observation (Fig.~\ref{fig:XLFs_merged}).  Because of non-uniform coverage in the
disk fields and because we test for variability from observation to observation, we carefully
selected sources in the field of view of all 16 observations (see Figs.~\ref{fig:outline} and
\ref{fig:src_coverage}).  As noted in \S~\ref{section:HR}, we do not remove contaminating
foreground/background sources in our fields on a source-by-source basis, since we cannot
differentiate such sources based on the X-ray observations alone.  We only make a global correction
to the XLFs for the expected AGN luminosity function, which we detail below.  Detailed source
matching, which will be completed in a future publication, will help to mitigate the contamination
of foreground/background sources and enable us to separate populations of sources in more detail
(Zezas et al. in prep.).

\subsection{Incompleteness Correction}
\label{section:XLFs--incompleteness}

The source detection threshold is not consistent over the entire field of view of each of the
observations or the merged observation because of a variety of instrumental and statistical effects
(e.g., varying PSF size and shape, effective detector area, etc.).  This leads us to detect only a
fraction of sources at lower luminosities, and, as a result, any XLF that we construct near the
detection threshold of our observations will be shallower due to incompleteness.  Therefore, we
correct all of our XLFs for incompleteness using the methods in \cite{zezas07} before any
comparisons and fits are attempted.

In brief, the detection probability of a source is calculated for multiple background levels and
off-axis angles from grids of MARX simulations \citep{wise03}.  Then, on a source-by-source basis,
we interpolate between these grids to estimate the detection probability and, hence, the
incompleteness correction factor to be applied to each source in each XLF.

We apply incompleteness corrections for each XLF based on one of two sets of cutoffs, each set
having two cutoffs:  a source count cutoff and an off-axis angle cutoff.  A source count cutoff is
important because we do not correct for Eddington bias \citep{eddington13} and an off-axis angle
cutoff is important because the detection threshold grows considerably toward large off-axis angles.
Each source must to have either more than 8.25 source counts and an off-axis angle less than 7
arcminutes or more than 15 source counts and an off-axis angle less than 9.5 arcminutes, depending
on which region of the galaxy it resides.  Beyond 9.5 arcminutes, the incompleteness correction for
almost any luminosity included in our source list is exceedingly large.  At count cutoffs greater
than 15 counts, we reject too many sources to construct useful XLFs.

For all of the XLF comparisons and fitting, we always used the largest off-axis angle cutoff for the
entire galaxy, inner disk, and outer disk regions, and the smaller off-axis angle cutoff for the
bulge.  We took care to uniformly impose our off-axis angle and count limits across all
observations, so that if a source fails to meet our criteria in one observation, the same source was
rejected in all other observations.  This is very important so that when we compare our XLFs, any
differences in the XLFs are from source variability alone, not source rejection.  Finally, for the
XLF tests for variability in the next section, we only compared the XLFs above certain luminosities
where the completeness of the XLFs is greater than $\sim 20\%$ (Table~\ref{table:XLF_cutoffs}).

\subsection{Testing for XLF Variability}
\label{section:XLFs--variability}

A large percentage of the individual sources included in the XLFs exhibit variability:
58\% (entire galaxy), 36\% (bulge), 60\% (inner disk), 43\% (outer disk).  Can this individual
source variability impart significant varibility to the XLFs?  We directly test this hypothesis.

When comparing our corrected XLFs, we used a non-parametric statistical test, the Kolmogorov-Smirnov
(K-S) test \citep{kolmogorov41}.  We also arrived at the same conclusions as below for the K-S test
when testing the uncorrected XLFs with the Kruskall-Wallis (K-W) \citep{kruskal52} and Mann-Whitney
(M-W) tests \citep{mann47}.  Helpful detailed explanations of these and other similar statistical
tests can be found in \cite{siegel56}.

The K-S test is useful because it could be applied to the incompleteness-corrected XLFs.  We tested
our XLFs for consistency by making pairwise comparisons of the XLFs using a two-sided K-S test,
which indicates how likely it is that the two XLFs were drawn from the same parent distribution.
A K-S statistic near 0 indicates that two data sets were not drawn from the same parent
distribution.

\begin{figure*}[t]
\centering
\includegraphics[scale=0.37, trim = 35mm 10mm 5mm 20mm, clip]{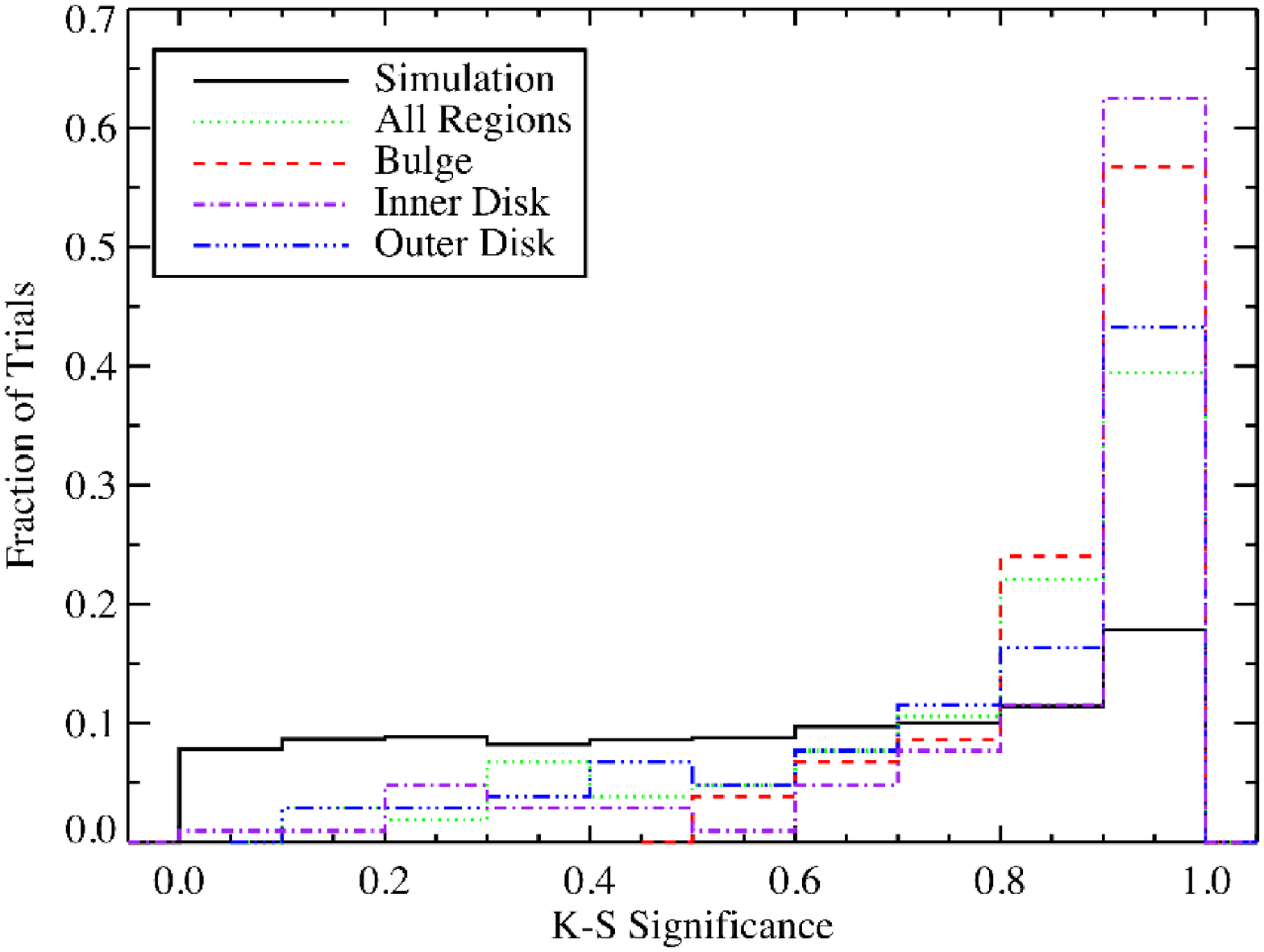}
\includegraphics[scale=0.37, trim = 35mm 10mm 5mm 20mm, clip]{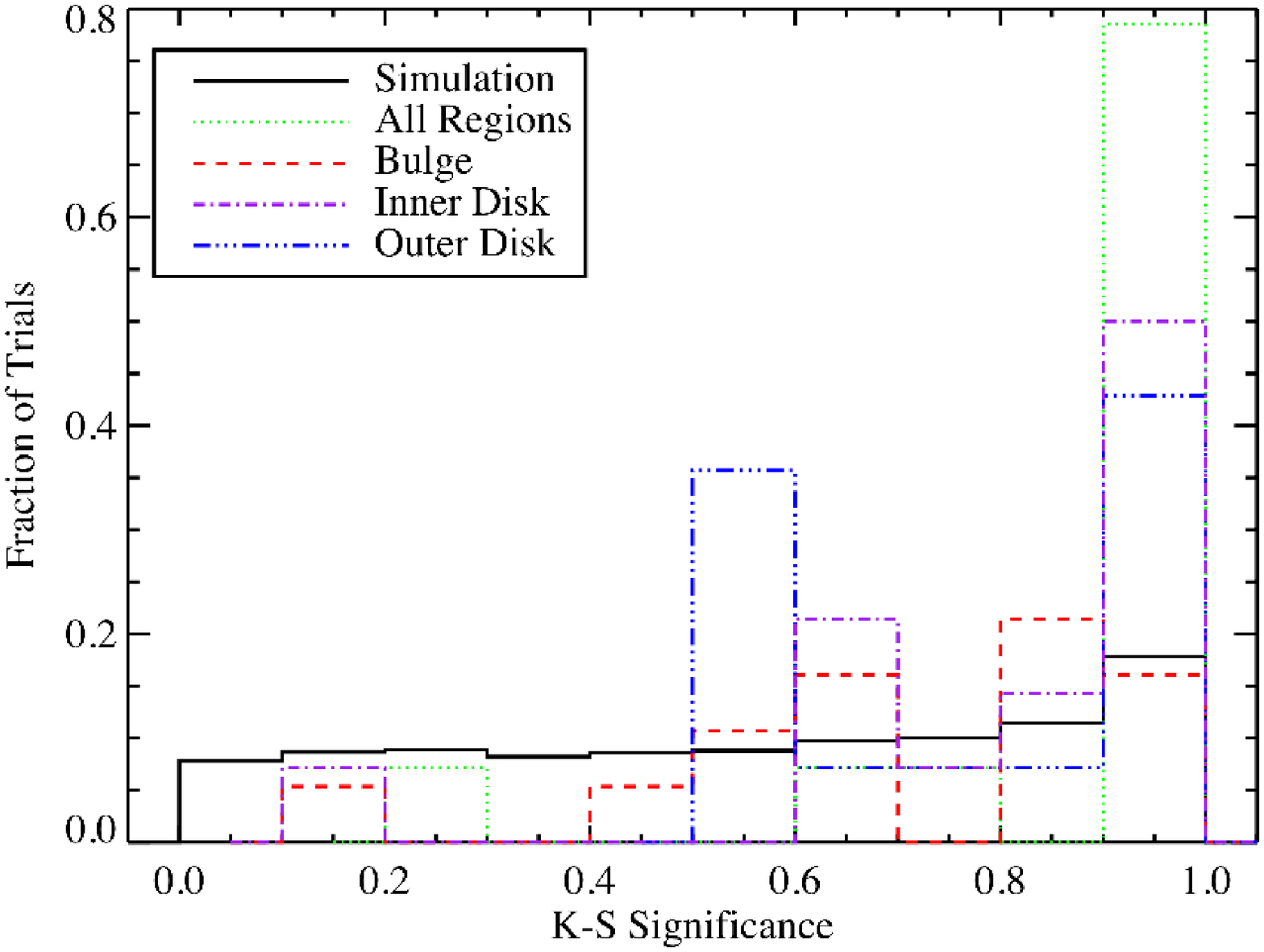}
\caption{The measured and simulated distribution of the K-S statistics between ObsIDs 5935--5949
(days--weeks timescale; left) and of ObsIDs 5935--5949 to ObsID 735 (5-year timescale; right).
The measured distributions are constructed from XLFs from sources in the field of view of all
observations included in the comparison, taking into account the off-axis angle, count, and
completeness thresholds discussed in \S~\ref{section:XLFs--incompleteness} and
\S~\ref{section:XLFs--variability}.
\label{fig:KS_dist}}
\end{figure*}

\begin{figure}[t]
\centering
\includegraphics[scale=0.38, trim = 30mm 20mm 20mm 20mm, clip]{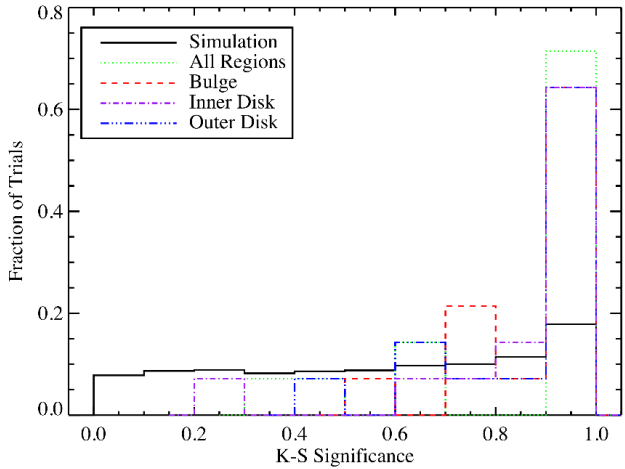}
\caption{The measured and simulated distribution of the K-S statistics for comparing each of the
individual ObsIDs 5935--5949 to their merged observation.  The measured distributions are
constructed from XLFs from sources in the field of view of all observations included in the
comparison, taking into account the off-axis angle, count, and completeness thresholds discussed in
\S~\ref{section:XLFs--incompleteness} and \S~\ref{section:XLFs--variability}.
\label{fig:KS_dist_merged}}
\end{figure}

To test the XLFs for consistency, we considered two different timescale comparisons separately, a
timescale of days--weeks (ObsIDs 5935--5949) and a timescale of $\sim 5$ years (ObsIDs 5935--5949 to
735 and merged to ObsID 735).  In doing so, we also compared the XLFs in each of our different
fields of view separately (entire galaxy, bulge, inner disk and outer disk regions).  In addition,
we test for the possibility that differences in the merged and individual XLFs arise because of
differences in individual source luminosities in a snapshot versus the average luminosity
\citep[see section 4.1 of][]{zezas07}.  Upon inspecting the distributions of these K-S significance
statistics (Figs.~\ref{fig:KS_dist} and \ref{fig:KS_dist_merged}), we find that, although most of
the K-S statistics are near 1 for all regions of the galaxy, some of them are near 0.

To explore the possible significance of this, we randomly generated 10000 XLFs by drawing 100
sources from a power-law distribution with index $-1.5$ (the exact number of sources and slope does
not matter).  Then, we performed the same pairwise K-S analysis on the generated XLFs and plotted
the distribution of K-S statistics (Figs.~\ref{fig:KS_dist} and \ref{fig:KS_dist_merged}).  Thi
Monte Carlo simulation shows that having a non-neglible number of K-S values near 0 is expected for
random samples drawn from the same parent distribution.  Futhermore, from the comparison of the
merged XLF to the ObsID 735 XLF, we found KS statistics:  0.98 (entire galaxy), 0.76 (bulge), 0.99
(inner disk) and 0.91 (outer disk).  Therefore, we conclude that the observed M81 XLFs for either
timescale are consistent with being drawn from the same distribution.  The intrinsic variability of
the individual M81 sources on these timescales does not make the XLFs inconsistent with each other,
suggesting that a snapshot survey provides a reliable indicator of the XLF.

\subsection{XLF Fitting}
\label{section:XLFs--fitting}

\begin{figure*}[tbp]
\centering
\includegraphics[scale=0.35, trim = 25mm 10mm 5mm 20mm, clip]{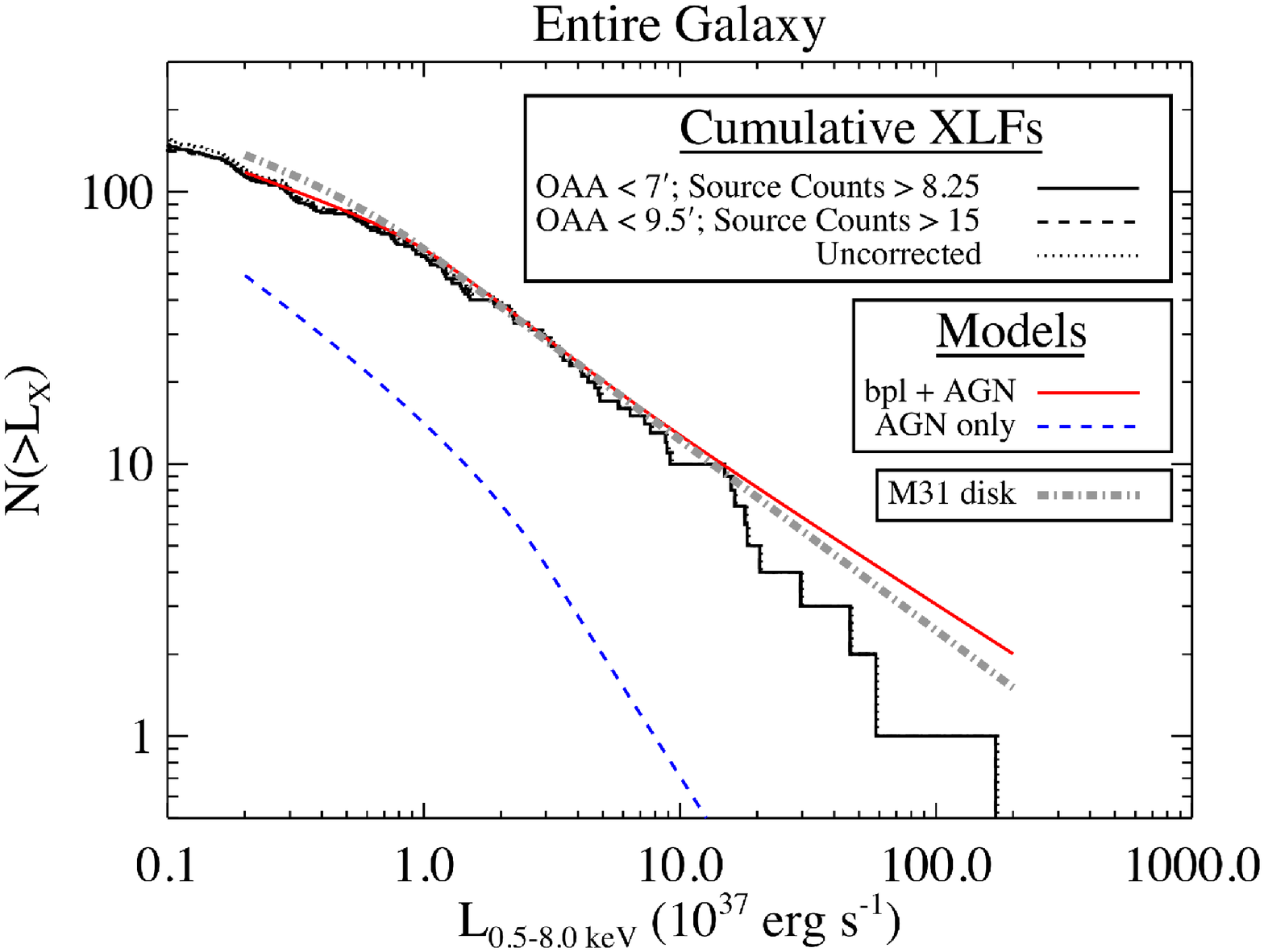}
\includegraphics[scale=0.35, trim = 25mm 10mm 5mm 20mm, clip]{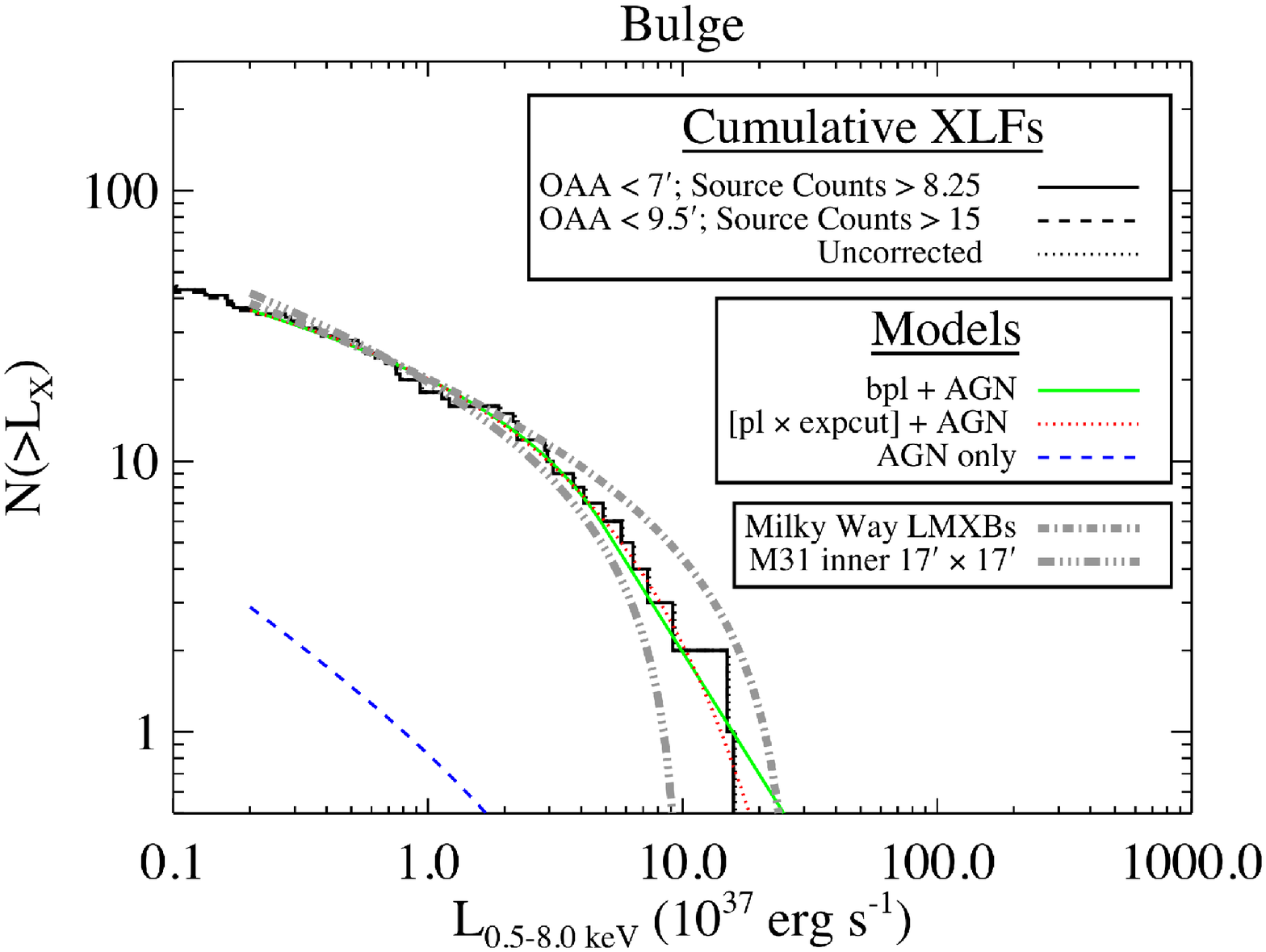}
\includegraphics[scale=0.35, trim = 25mm 10mm 5mm 20mm, clip]{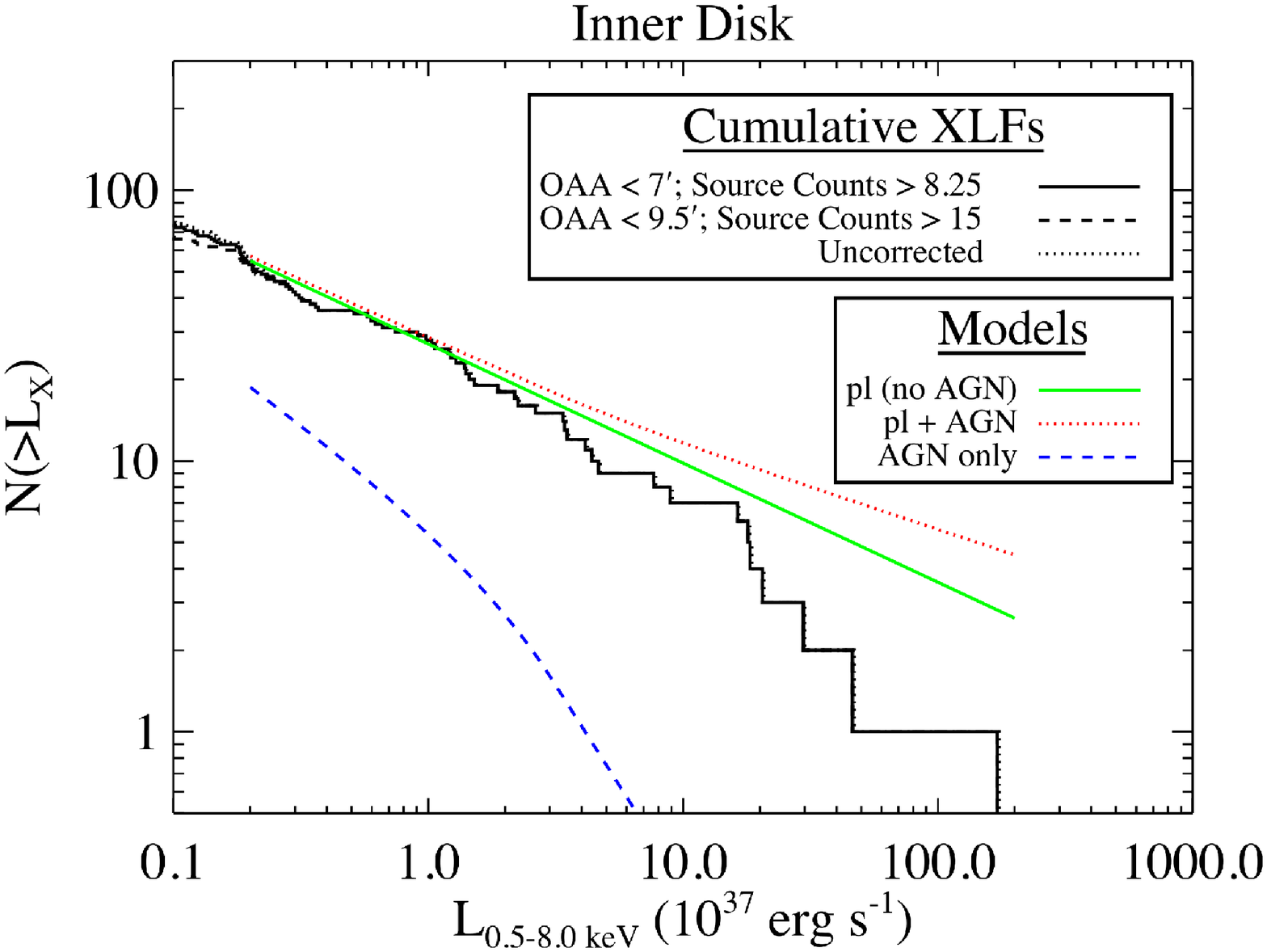}
\includegraphics[scale=0.35, trim = 25mm 10mm 5mm 20mm, clip]{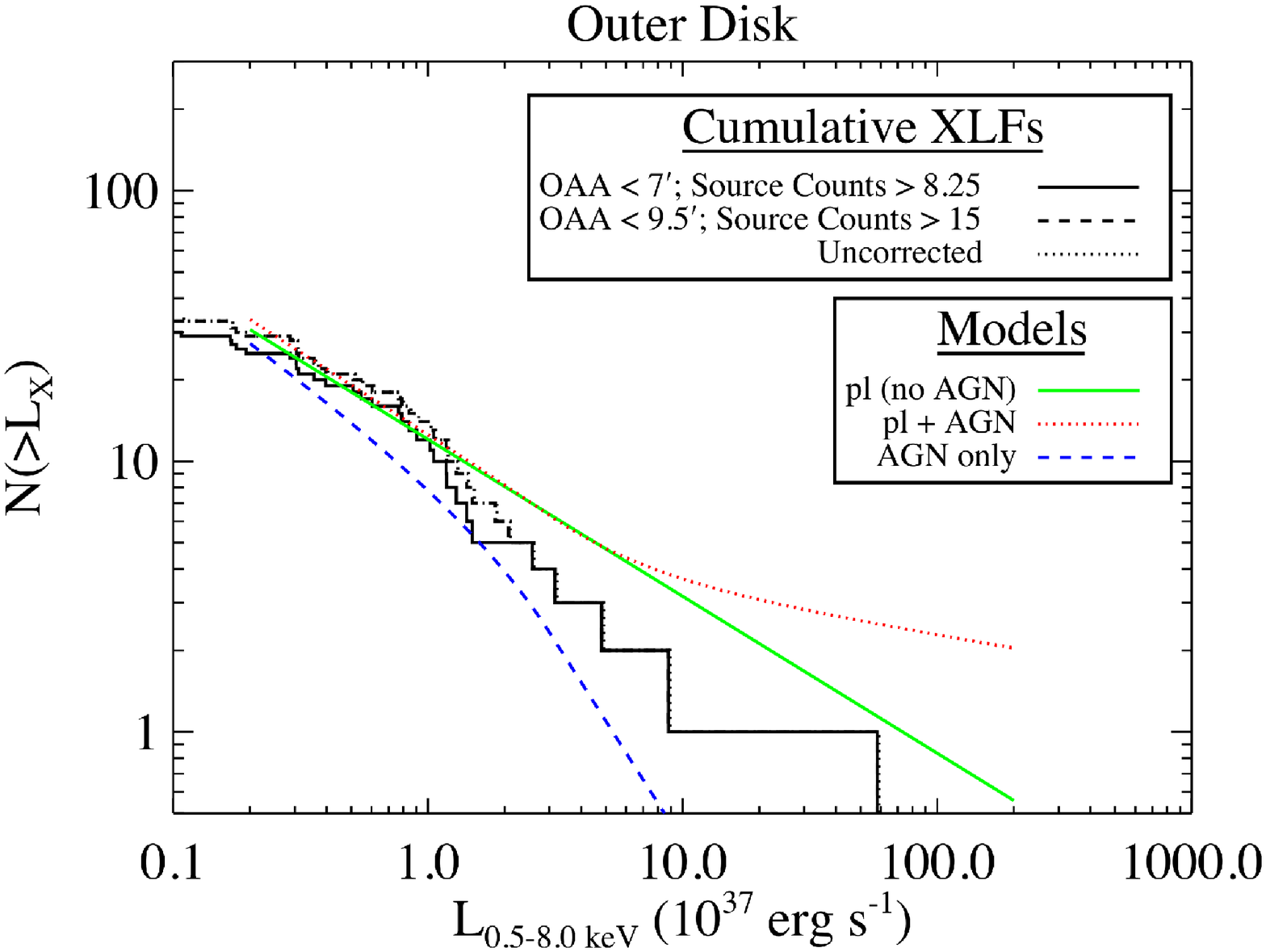}
\caption{Merged cumulative XLFs for sources that are in the field of view of the merged ObsIDs.
Each plot represents a different region.  The best-fit power-law models are plotted as listed in
Table~\ref{table:XLF_fits} (see \S~\ref{section:XLFs--fitting}).  We compare our entire galaxy XLF
to the total disk XLF of M31 \citep{shawgreening09} by normalizing it to the M81 models at
$10^{37}$~erg~s$^{-1}$ and plotting it as dim dash-dotted line on our entire galaxy XLF.  We also
compare our bulge XLF to the LMXBs for the Milky Way \citep{grimm02} and the inner
$17^\prime \times 17^\prime$ (bulge) of M31 \citep[the cutoff power-law model][]{kong03} as above.
Note that all of these studies use different energy ranges and methods for calculating the
luminosities, which contributes to systematic differences in the XLFs that are not accounted for
here.  Also, the XLF fits are affected by the contribution of AGN to varying degrees, depending on
the region of the galaxy considered.  The expected contribution of AGN to the low-luminosity end of
the XLFs, in particular, appears to be too large for the M81 field.  The correction for AGN and its
limitations are discussed in more detail in \S~\ref{section:XLFs--fitting}.
\label{fig:XLFs_merged}}
\end{figure*}

Parameterizing the XLFs is useful for comparing our XLFs to XLFs in other galaxies and those created
by synthesis models.  Therefore, we fit the differential XLFs with simple and modified power-law
functions\footnote{Even though the fit is visualized on cumulative luminosity functions, the fits
are always performed on the differential number of sources.}.  There are a number of methods that
can be used to fit a differential number of points in an unbiased way:  maximum likelihood methods
(\citealt{clauset07}, \citealt{zezas07}), a method using the K-S test \citep{johnston96}, or methods
using X-ray spectral fitting software (\citealt{kenter03}, \citealt{zezas07}).

\subsubsection{Our XLF Fitting Method and Models}
\label{section:XLFs--method_models}

We choose to use the method utilizing X-ray spectral fitting software because it is convenient and
has been shown to yield consistent results with a maximum likelihood method \citep[see][]{zezas07}.
Furthermore, we can implement the incompleteness correction through an ARF that scales the
differential number of sources for the fitting.

Because of our large dynamic range for fitting
($2 \times 10^{36}$--$2 \times 10^{39}$~erg~s$^{-1}$), we rebin the XLF and ARF in counts space to
six-count bins as opposed to a ``natural" binning scheme of one-count bins.  This prevents the ARFs
from becoming too large, which results in the fits taking a very long time to run or Sherpa
crashing.  Over a smaller dynamic range, we verified that the best-fit parameters do not change
significantly when we use this variation of the ``natural" binning scheme.

Both the ARF and XLF are read into Sherpa 3.4 and fit just like a typical X-ray spectrum with the
C-Statistic since the number of counts in each bin are very small.  We first tried fitting the
differential merged XLFs for each of the four regions (entire galaxy, bulge, inner disk, outer disk)
with a single power law of the form:

\begin{equation}
\label{eq:1pl}
\frac {dN} {dL} = A \left ( \frac {L} {L_{ref}} \right )^{- \alpha_1}
\end{equation}
where the reference luminosity, L$_{\text{ref}}$, is always set to $10^{37}$~erg~s$^{-1}$ and A is
the amplitude at the reference luminosity (cumulative slope = $\alpha_1 - 1$).  We also considered a
broken power law of the form:
\begin{equation}
\label{eq:2pl}
\frac {dN} {dL} = \left \{ \begin{matrix} A \left( \dfrac {L} {L_b} \right)^{-\alpha_1}
	\quad \text{for} \ L \le L_b \\ A \left( \dfrac {L} {L_b} \right)^{-\alpha_2}
	\quad \text{for} \ L \ge L_b \end{matrix} \right .
\end{equation}
so that the amplitudes and reference luminosities of the two power laws match and the reference
luminosity is defined as the break luminosity, L$_\mathrm{b}$.  In addition, given that the bulge
XLF appears truncated, we also fit a power law with an exponential cutoff to it of the form:
\begin{equation}
\label{eq:pl_exp}
\frac {dN} {dL} = A \left ( \frac {L} {L_{ref}} \right )^{- \alpha_1} e^{ C (L - L_{b}) }
\end{equation}
where C is a constant and L$_\mathrm{b}$ is the offset or break luminosity.

Lastly, AGN number counts (\citealt{rosati02}, \citealt{bauer04}, and \citealt{kim07}) suggest that
background AGN make up a non-negligible fraction of the sources in our XLFs.  We attempt to
compensate for the AGN contribution in our model fits by adding a fixed flux distribution of AGN to
all of the models in equations~\eqref{eq:1pl}--\eqref{eq:pl_exp} before fitting.  As is clear from
work on the number counts of AGN mentioned above, we can describe the distribution by a broken power
law with differential indices $\sim 1.6$ and 2.5 (Euclidean) below and above the break.  We use the
broken power-law model (different in functional form to ours above) and fit parameters exactly as
given in the number count results of the CHAMP survey
\citep[see equations (3)-(5) and Table 3 of][]{kim07}.  We use the row of parameters in Table 3 of
\citealt{kim07} derived using a photon index of 1.7 and that match our energy range (0.5--8.0 keV).

We report all of our fit results excluding and including the AGN contribution in
Table~\ref{table:XLF_fits}.

\begin{center}
\begin{deluxetable*}{lcccccccc}
\tablecaption{Merged XLF Fits}
\tablehead{
	\colhead{} &
	\multicolumn{4}{c}{Without the AGN Model} &
	\multicolumn{4}{c}{With the AGN Model} \\
	\colhead{} &
	\colhead{$\alpha_1$} &
	\colhead{$\alpha_2$ or} &
	\colhead{Amplitude$^\dagger$} &
	\colhead{L$_\mathrm{b}$} &
	\colhead{$\alpha_1$} &
	\colhead{$\alpha_2$ or} &
	\colhead{Amplitude$^\dagger$} &
	\colhead{L$_\mathrm{b}$} \\
	\colhead{Region (Model)} &
	\colhead{} &
	\colhead{Coeff.$^\ddagger$} &
	\colhead{} &
	\colhead{($10^{37}$ erg s$^{-1}$)} &
	\colhead{} &
	\colhead{Coeff.$^\ddagger$} &
	\colhead{} &
	\colhead{($10^{37}$ erg s$^{-1}$)} }
\startdata
All & 1.00${_{-0.19}^{+0.12}}$ & 1.67${_{-0.08}^{+0.10}}$ & 29.43${_{-3.44}^{+3.32}}$ &
	1.21${_{-0.09}^{+0.23}}$ & 0.26${_{-0.60}^{+0.35}}$ & 1.60${_{-0.08}^{+0.12}}$ &
	21.49${_{-3.36}^{+4.50}}$ & 1.20${_{-0.22}^{+0.14}}$ \\
Bulge (bpl) & 1.06${_{-0.18}^{+0.09}}$ &
	2.49${_{-0.40}^{+0.54}}$ & 1.77${_{-0.44}^{+0.32}}$ & 4.87${_{\mathrm{NA}}^{\mathrm{NA}}}$ &
	1.01${_{-0.20}^{+0.11}}$ & 2.50${_{-0.41}^{+0.56}}$ & 1.75${_{-0.45}^{+0.35}}$ &
	4.87${_{\mathrm{NA}}^{\mathrm{NA}}}$ \\
Bulge (pl $\times$ expcut) & 0.92${_{-0.25}^{+0.23}}$ & -0.12${_{-0.08}^{+0.06}}$ &
	4.20${_{-3.80}^{\mathrm{NA}}}$ & 8.00${_{\mathrm{NA}}^{\mathrm{NA}}}$ & 0.86${_{-0.27}^{+0.25}}$ &
	-0.13${_{-0.08}^{+0.06}}$ & 2.87${_{-2.82}^{\mathrm{NA}}}$ & 9.90${_{\mathrm{NA}}^{\mathrm{NA}}}$ \\
Inner Disk & 1.44${_{-0.08}^{+0.09}}$ & NA & 11.91${_{-1.59}^{+1.75}}$ & NA &
	1.31${_{-0.11}^{+0.11}}$ & NA & 7.21${_{-1.68}^{+1.82}}$ & NA \\
Outer Disk & 1.58${_{-0.12}^{+0.13}}$ & NA & 6.99${_{-1.22}^{+1.38}}$ & NA &
	1.16${_{-1.07}^{+0.37}}$ & NA & 0.76${_{-0.84}^{+1.32}}$ & NA \\
\enddata
\tablecomments{Best XLF model fit parameters as defined in equations
\eqref{eq:1pl}--\eqref{eq:pl_exp} with and without the background AGN broken power-law
model contribution (based on AGN number counts).  We fit and report the fit information for the
differential XLFs.  All of the XLFs were fit over the range
$2 \times 10^{36}$--$2 \times 10^{39}$~erg~s$^{-1}$, where the completeness is greater than
$\sim 50\%$.  The slopes are systematically shallower with the AGN model included and shallower than
expected compared to XLF slopes of similar galaxies.  This suggests that the expected contribution
of AGN to the low-luminosity end of the XLFs, in particular, could be too large for the M81 field.
See \S~\ref{section:XLFs--fitting} for model definitions and further background AGN discussion.  In
various places throughout the table, NA = ``Not Applicable/Available" is used because there is
either no slope, coefficient, or break luminosity for a straight powerlaw or because the Sherpa
uncertainty estimates did not converge. \\
($\dagger$) The power-law reference points for the amplitude (number of sources at the
reference luminosity in the differential XLF) are always equal to $10^{37}$ erg s$^{-1}$ for the
single power-law model or the exponential cutoff model for the bulge and equal to the break
luminosity for the broken power-law model. \\
($\ddagger$) In the case of the bulge power law with an exponential cutoff, the exponential
coefficient, ``C", as in equation~\eqref{eq:pl_exp}.}
\label{table:XLF_fits}
\end{deluxetable*}
\end{center}

\subsubsection{Discussion of XLF Fit Results by Region}
\label{section:XLFs--fit_dis}

First, the contribution from AGN is a serious issue in most of our fields.  The AGN contribution
increases at smaller fluxes because it appears that the LogN--LogS slope is steeper than the slope
of the M81 XLF.  At face value, including the AGN distribution forces the best-fit power-law slopes
for M81 to be shallower in all of the XLFs.  However, foreground absorption of M81 brings about an
uncertainty in the AGN flux distribution.  The foreground absorption is difficult to quantify
because the clumpiness of the disk suggests a highly variable column density.  A disk scale height
of a few hundred parsecs for a typical ISM density of $\sim 1$~cm$^{-3}$ at the inclination of M81
\citep[$\sim 35^{\circ}$;][]{boggess59} produces an average column density $\sim 10^{21}$~cm$^{-2}$
measured perpendicular to the disk.  This could decrease the flux of a typical AGN by $\sim 10\%$,
flattening the faint-end slope of logN--LogS for AGN in this region.

The uncertainty from foreground absorption together with galactic foreground diffuse emission from
M81, standard errors in the survey measurements, and cosmic variance make interpreting our XLFs very
difficult, especially at the faint ends.  For example, an uncertainty of $\sim 20\%$ due
to cosmic variance and an equally sizable shift brought about by the foreground emission and
absorption of M81 changes the best-fit slope of the outer disk by $\sim 0.1$ and the amplitude by
$\sim 70\%$.  Note that the AGN contamination varies with the galactic source density (see
Table~\ref{table:XLF_fits} and Figure \ref{fig:XLFs_merged}) so that the bulge XLF is least affected
and the outer disk XLF is most affected.  In light of these complications, we are still able to make
some concrete statements regarding our XLFs.

In order to interpret our XLF fits, we need to compare the significance of the fits between the
single and broken power-law models.  Since the C-Statistic, a maximum likelihood statistic, is used,
it is appropriate to compare the quality of the fits using a likelihood ratio test
\citep[e.g.,][]{zezas07}.  We simulate 1000 XLFs from a single power-law model and then fit them
with the single and the broken power-law model in the same way that we fit the observed XLFs.  We
then calculate the ratio of the single to the broken power-law best-fit C-statistics for each XLF. 
The confidence level corresponding to the amount of improvement of the fit from the single to the
broken power-law model is the fraction of times that the simulated ratio of statistics is greater
than the measured ratio of the statistics.  One should not judge the quality of the fits from the
best-fit cumulative distribution functions on the logarithmically-scaled plots because they can be
misleading.  For instance, uncertainties in the XLFs at the high-luminosity end are much larger than
those at the low-lumionsity end.  In addition, unexpected statistical effects frequently caused by
the skewness in the probability distribution of sources comprising the XLFs have been documented
previously (\citealt{gilfanov04b}; \citealt{clauset07}).

In the bulge, we find that the broken power-law model and the power-law model with an exponential
cutoff provide a highly-significant improvement in the fit versus a single power law ($> 99.9\%$
confidence, $\gg 3 \sigma$).  The break luminosity is poorly constrained but is within a factor of a
few of the Eddington luminosity of a neutron star and is consistent with the values derived from
other previous work for elliptical and S0 galaxies, the bulges of other galaxies, and the LMXBs of
our Galaxy \citep{sarazin00, blanton01, kundu02, grimm02, kim06, voss09}.  We also see evidence of
the flattening of the LMXB XLF below $\sim 10^{37}$~erg~s$^{-1}$ that has been seen by many of
these previous studies.  The best fit functions with the AGN model are
\begin{equation}
\label{eq:2pl_bestfit_bulge}
\frac {dN} {dL} = \left \{ \begin{matrix} 1.75 \left( \dfrac {L} {L_b} \right)^{-1.01}
	\quad \text{for} \ L \le L_b \\ 1.75 \left( \dfrac {L} {L_b} \right)^{-2.50}
	\quad \text{for} \ L \ge L_b \end{matrix} \right .
\end{equation}
for the broken power law, where L$_\text{b} = 4.87 \times 10^{37}~\mathrm{erg~s}^{-1}$ and
\begin{equation}
\label{eq:pl_exp_bestfit}
\frac {dN} {dL} = 2.87 \left(\frac{L}{10^{37}\,{\rm erg\,s^{-1}}}\right)^{-0.86}e^{-0.13\left( L - L_b \right)}
\end{equation}
for the power law with an exponential cutoff, where
L$_\text{b} = 9.9 \times 10^{37}~\mathrm{erg~s}^{-1}$.

The shape of this XLF together with the locations of the sources in the color-color plot
(Fig.~\ref{fig:HR}) suggests a very old population of stars dominates the innermost part of M81 and
that we are probing a population of mostly LMXBs.  The shape of the XLF is also consistent with the
overall shape of the average LMXB XLF, which has a flat cumulative distribution below a few times
$10^{37}$~erg~s$^{-1}$ with a cutoff near a few times $10^{38}$~erg~s$^{-1}$ \citep{gilfanov04a}.
The AGN do not strongly bias these results as they are expected to comprise $\lesssim 10\%$ of the
sources in the XLF above the cutoff luminosity, $2 \times 10^{36}$~erg~s$^{-1}$.

Next, because of the very large percentage of AGN expected in the outer disk XLF ($\sim 80\%$ of the
sources above the cutoff luminosity and $\sim 60\%$ of the sources above $10^{37}$~erg~s$^{-1}$),
this region is the most difficult to interpret.  This region also has the fewest total number of
sources, and brief inspection of some all-sky optical surveys suggests that there are also a few
foreground stars, which we have not attempted to remove.  While there appears to be a break near
$10^{37}$~erg~s$^{-1}$, it does not bring about a significant improvement in the fit when the AGN
contribution is taken into account (94\% confidence, $1.9 \sigma$).

Therefore, we fit this XLF with a single, unbroken power law as is typically seen in disk-like
regions of ongoing star formation as in the disk XLF of our Galaxy \citep{grimm02}, the Antennae
\citep{zezas07}, and NGC 6946 \citep{fridriksson08}, for examples.  The best-fit function with the
AGN model is
\begin{equation}
\label{eq:1pl_bestfit_OD_wAGN}
\frac {dN} {dL} = 0.76 \left ( \frac {L} {10^{37}~\mathrm{erg~s}^{-1}} \right )^{-1.16}
\end{equation}
with a much shallower slope than what has been found in the studies above and which produces an
unusual-looking fit (Fig.~\ref{fig:XLFs_merged}).  However, the best-fit function without the AGN
model is
\begin{equation}
\label{eq:1pl_bestfit_OD_woAGN}
\frac {dN} {dL} = 6.99 \left ( \frac {L} {10^{37}~\mathrm{erg~s}^{-1}} \right )^{-1.58}
\end{equation}
with a slope that is more consistent with the other disk studies above.  The slope cannot be
well-constrained and, given the large fit uncertainties (the errors for the slopes are on the order
of the range of slopes found with or without the AGN model), we do not attempt to interpret this
XLF further.

Our inner disk region XLF is consistent with a single, unbroken power law and a broken power law
does not result in a significant improvement in the fit (44\% confidence, $< 1 \sigma$).  The
best-fit function with the AGN model is
\begin{equation}
\label{eq:1pl_bestfit_ID_wAGN}
\frac {dN} {dL} = 7.21 \left ( \frac {L} {10^{37}~\mathrm{erg~s}^{-1}} \right )^{-1.31}
\end{equation}
Inspection of the color-color plot (Fig.~\ref{fig:HR}) indicates that there are a population of
LMXBs in this region (or HMXBs with black holes that have very similar colors) that are embedded
with the HMXBs and other sources (this is also where the ULXs and a few SNRs are).  While visual
inspection of Figure~\ref{fig:XLFs_merged} may suggest that this XLF has a break, the fits indicate
that it is not significant.  Furthermore, the best-fit power-law slope and the fact that
well-pronounced spiral arms are present, combined with an inspection of Fig.~\ref{fig:HR}, supports
the notion that this XLF appears very disk-like and that a large fraction of the sources in this
region are HMXBs.  The AGN make a moderate contribution to the sources in the XLF here ($\sim 30\%$
of the sources above the cutoff luminosity) and do change the slope and normalization of the
best-fit model when taken into account.  However, whether we take into account the AGN contribution
or not, a single power law is strongly favored and the best-fit slopes in either case are consistent
with each other.

A global view of X-ray point source population is revealed in the total XLF.  There is a marginal
improvement in the fit from the single to the broken power law (97\% confidence, $2.2 \sigma$) with
a break near $10^{37}$~erg~s$^{-1}$, which could be the break in the bulge manifesting itself at a
slightly lower luminosity than measured in the bulge.  The best-fit function with the AGN model is
\begin{equation}
\label{eq:2pl_bestfit_all_wAGN}
\frac {dN} {dL} = \left \{ \begin{matrix} 21.49 \left( \dfrac {L} {L_b} \right)^{-0.26}
	\quad \text{for} \ L \le L_b \\ 21.49 \left( \dfrac {L} {L_b} \right)^{-1.60}
	\quad \text{for} \ L \ge L_b \end{matrix} \right .
\end{equation}
where L$_\text{b} = 1.20 \times 10^{37}~\mathrm{erg~s}^{-1}$ and without the AGN model is
\begin{equation}
\label{eq:2pl_bestfit_all_woAGN}
\frac {dN} {dL} = \left \{ \begin{matrix} 29.43 \left( \dfrac {L} {L_b} \right)^{-1.00}
	\quad \text{for} \ L \le L_b \\ 29.43 \left( \dfrac {L} {L_b} \right)^{-1.67}
	\quad \text{for} \ L \ge L_b \end{matrix} \right .
\end{equation}
where L$_\text{b} = 1.20 \times 10^{37}~\mathrm{erg~s}^{-1}$.
The slopes with or without the AGN model are fairly shallow and resemble the XLF slope of the inner
disk region of M81 and disk regions in other galaxies where there is considerable star formation
occurring.  However, because this XLF does also have a significant break and the color-color plot
(Fig.~\ref{fig:HR}) suggests contributions from many different populations of sources, we conclude
that this XLF is neither disk- nor bulge-dominated, but is a fairly even mixture of both types.

Finally, we can use our XLFs to estimate two interesting galactic properties:  the stellar mass and
the SFR, using relationships in \cite{gilfanov04a} and \cite{grimm03},
respectively.

First, to estimate the total stellar mass of M81, we need to estimate the number of LMXBs in M81
with L$_\text{X} > 10^{37}$~erg~s$^{-1}$.  Since we have found that the bulge is likely
almost exclusively composed of LMXBs, we use our bulge XLF to estimate that there are $\sim 20$
LMXBs in this range.  As we have stated above, we also expect there to be a population of LMXBs in
the disk regions as well, especially in the inner disk region.  The fraction of LMXBs can be
estimated from our color-color plot (Fig.~\ref{fig:HR}) but is very uncertain.  Taking this fraction
and applying it to our XLF in this luminosity range, we estimate that there are $\sim 10$ LMXBs in
this region.  The outer disk region appears to contribute only a very small amount of LMXBs with
L$_\text{X} > 10^{37}$~erg~s$^{-1}$ when the expected AGN fraction is taken into account, so
we do not include it here.

Then using these numbers together with equation 20 in \cite{gilfanov04a}, we estimate that the
total stellar mass of M81 is $> 2.1 \times 10^{10}$~M$_\sun$.  This is a fairly robust lower limit
to the number LMXBs in M81 because:  we have neglected the outer disk contribution, the K-band light
is measured out to larger radii than our X-ray data, there are some sources superimposed against the
LINER nucleus that we are unable to detect or classify (some of these are the ML sources in
Table~\ref{table:ML_sources}), and equation 21 suggests a larger mass.

We also estimate the total stellar mass of M81 using the total 2MASS K-band magnitude,
$\sim 3.9$~mag.  This corresponds to a solar K-band luminosity of $7.7 \times 10^{10}$~L$_\sun$.
Using M$_*$/L$_\text{K} \approx 1.1$ from GALEV models for an SAab-type galaxy
\citep{kotulla09}, we estimate a total stellar mass of $8.5 \times 10^{10}$~M$_\sun$.  Reversing the
calculation, this stellar mass suggests that M81 has $\sim 70$--120 LMXBs with
L$_\text{X} > 10^{37}$~erg~s$^{-1}$.  Part of this discrepancy can be attributed to the reasons
listed at the end of the previous paragraph.  Other reasons include uncertainties in the
M$_*$/L$_\text{K}$ ratio of M81 and of the values assumed by \cite{gilfanov04a}, which, together,
could contribute deviations of factors of approximately two to three.  For these reasons, caution
should be taken in estimating the number of LMXBs and stellar mass using this method for M81 and
other galaxies with similar issues.

Second, to estimate the SFR, we need to estimate the total number of HMXBs and their total
luminosity for L$_\text{2--10 keV} > 2 \times 10^{38}$~erg~s$^{-1}$.  There are four sources with
L$_\text{2--10 keV} > 2 \times 10^{38}$~erg~s$^{-1}$.  Inspection of the locations of these sources
in Fig.~\ref{fig:HR} suggests that at least one of these sources is likely to be a HMXB.  The other
three sources all lie on the boundary between HMXBs and LMXBs.  This, combined with the knowledge
that the color of an individual source is not sufficient to determine the precise nature of the
source (\S~\ref{section:HR}), and the fact that we are subject to large Poisson errors at
the high-luminosity end of the XLF, only allows us to place approximate limits on the SFR estimated
with this method.  Equation 20 of \cite{grimm03} suggests a range of SFRs of
0.34--1.38~M$_\sun$~yr$^{-1}$ for 1--4 sources with
L$_\text{2--10 keV} > 2 \times 10^{38}$~erg~s$^{-1}$, and equation 22 of \cite{grimm03} suggests a
range of SFRs of 0.52--1.02~M$_\sun$~yr$^{-1}$ for the range of luminosities of these sources,
1.34--$2.64 \times 10^{39}$~erg~s$^{-1}$.  These estimates are comparable to the SFR that has been
estimated for M81 based on a combination of ultraviolet, H$\alpha$, and infrared
observations \cite[0.3--0.9~M$_\sun$ yr$^{-1}$; e.g.,][]{gordon04}.

\section{Comparing the XLFs of M81 with Those of Two Other Nearby Similar Spiral Galaxies}
\label{section:general_comps}

M81, M31, and our Milky Way are all Sb-type galaxies with similar SFRs near 1~M$_\sun$~yr$^{-1}$. 
Below, we make comparisons of the global properties of their X-ray binary populations using their
XLFs.

\subsection{Comparisons With the Milky Way}
\label{section:comp_M81_MW}

First, the XLFs of both the inner and outer disk regions of M81 are systematically shallower than
the HMXB slope of the Milky Way, but, given the statistical uncertainties and uncertainties in the
AGN contribution, are fairly consistent with one another.  While the M81 inner and outer disk
regions do appear to be dominated by HMXBs, they also contain other types of sources (e.g., AGN,
SNRs) that we are unable to separate conclusively on an individual basis at the present time.

Second, the near match of the M81 bulge and the normalized\footnote{Given the biases in determining
matching fields of view, blended and hidden sources in the bulge of M81, AGN contributions, etc., we
do not compare source densities in our regions.} LMXB population in the Milky Way is striking
(Fig.~\ref{fig:XLFs_merged}).  We find that the low-luminosity LMXB slope of the Milky Way is
similar to the low-luminosity slope found for the M81 bulge within the uncertainties.  Also, the
cutoffs in the LMXB XLFs are within a factor of $\sim 2$ of one another, supporting the notion that
the bulge of M81 is dominated by LMXBs as seen in the color-color plot (Fig.~\ref{fig:HR}).

\subsection{Comparisons With M31}
\label{section:comp_M81_M31}

We find that the normalized XLF of the inner $17^\prime \times 17^\prime$ region of M31 as measured
with Chandra by \cite{kong03} is quite similar to the bulge XLF of M81 (Fig.~\ref{fig:XLFs_merged}).
These data also match well to the bulge region observations measured with XMM-Newton
(r$ < 15^\prime $) by \cite{trudolyubov02}.

The disk XLFs of M31 are more difficult to compare as studies of the M31 disk return different
results (\citealt{kong03}, \citealt{shawgreening09}).  Although the XLFs in each of the disk regions
are consistent with single, unbroken power law, as expected, the power-law slopes measured by each
group are significantly different.  We examine some of the reasons for these differences below.

The disk XLFs of \cite{kong03} do not include ULXs or any sources brighter than $\sim 3 \times
10^{37}$~erg~s$^{-1}$ (unlike the disk XLFs in M81 and the HMXB XLF of the Milky Way).  However, the
field of view of Chandra is much smaller than M31 and all of the Chandra fields in \cite{kong03}
together still only cover a fraction of the disk of M31.  As noted in \cite{shawgreening09}, there
are much brighter sources seen in their fields, which cover most of the M31 disk that are being
missed in small-area surveys.  Missing these brighter sources could truncate the high-luminosity end
of \cite{kong03}'s disk XLFs, and could make these XLFs steeper.  However, the effect that a few
bright sources have on the best-fit slope is not obvious without further detailed analysis, which is
beyond the scope of this paper.

Second, \cite[as pointed out in][see their Figure 5]{shawgreening09}, model bias in the spectral
fitting of each of the point sources that comprise the XLF could affect the best-fit normalization
and slope of the XLF.  For instance, \cite{kong03} fit all of their sources with a fixed, absorbed
power law, which could be biasing their XLFs.  We could also be slightly biased by choosing a
power law as our initial fitting model, but we expect this to have a smaller effect because we have
allowed for the possibility of a model that better describes softer sources (blackbody).  In
addition, we float our photon index or blackbody temperature for all sources with more than 25
counts and always float our column densities for all sources with more than 5 counts.

Thoroughly assessing the impact of a single versus a variable conversion factor on the XLFs directly
is beyond the scope of this paper.  However, we can make some simple statistical comparisons for the
merged luminosities of the sources in the master source list.  We compare the error that would arise
from using a single conversion factor to calculate the luminosities (Fig.~\ref{fig:conv_factor}) and
the statistical error in the luminosity scaled from the Bayesian error in the counts.  We find that
the shape of both distributions is very similar, but that the median and the width of the
distribution of the statistical percent error is greater than the median (by approximately a factor
of 1.5) and the width of the distribution of the percent error that would arise from using a single
conversion factor.  Because the conversion factor errors are typically smaller than the statistical
errors in the luminosity, we suspect that the effect on the XLFs would probably be fairly small, but
maybe non-negligible.

In light of these issues, we only compare our disk and entire galaxy XLF results to the disk XLFs
constructed by \cite{shawgreening09}.   The total disk XLF that these authors construct surprisingly
seems to better match the M81 XLF of the entire galaxy rather than the inner or outer disk XLFs.
First, they claim a break in their total disk XLF, which happens to coincide with the measured break
in the M81 XLF of the entire galaxy.  Also, the power-law slopes are in good agreement with the
slopes that we measure for M81's entire galaxy XLF.  These results suggest that the disk of M31 has
a high-luminosity component similar to what is seen in the inner and outer disks of M81, but it also
seems to have a more considerable population of LMXBs there that produce at least a marginally
significant break near $10^{37}$~erg~s$^{-1}$.

\section{Summary}
\label{section:summary}

We have presented an in-depth analysis of multiple aspects of the X-ray source population of M81
from the perspective of the X-ray observations alone.  We conducted our source detection procedure
carefully, producing a master source list of 265 sources.  We then extracted and individually fit
the spectra for each of these sources in each of the 16 observations of M81 included in our
analysis.  With the spectral information, we calculated hardness ratios and used the aid of a
color-color plot to classify different populations of sources.  From the population study, we
devised an alternate conservative method of separating the bulge and the disk of the galaxy.  Then,
we quantified the variability of the individual sources and considered their possible effect on the
luminosity function of the galaxy.  We find that, despite detecting significant variability in
$\sim 36$--60\% of the sources included in the XLFs, the XLF of M81 remains stable at luminosities
greater than $\sim 2 \times 10^{37}$~erg~s$^{-1}$.  Finally, we plotted and fit the XLFs, analyzed
them with regard to their point source populations, and compared them to the XLFs of M31 and the
Milky Way.

\acknowledgments
We thank Jay Gallagher and Christy Tremonti for helpful comments.  DP and PS gratefully acknowledge
support from Chandra grant GO5-6092.  AZ acknowledges support by the UE IRG grant 224878.  Space
Astrophysics at the University of Crete is supported by EU FP7-REGPOT grant 206469 (ASTROSPACE).

\bibliographystyle{apj}

\begin{thebibliography}{72}
\expandafter\ifx\csname natexlab\endcsname\relax\def\natexlab#1{#1}\fi

\bibitem[{{Angelini} {et~al.}(2001){Angelini}, {Loewenstein}, \&
  {Mushotzky}}]{angelini01}
{Angelini}, L., {Loewenstein}, M., \& {Mushotzky}, R.~F. 2001, \apjl, 557, L35

\bibitem[{{Bauer} {et~al.}(2004){Bauer}, {Alexander}, {Brandt}, {Schneider},
  {Treister}, {Hornschemeier}, \& {Garmire}}]{bauer04}
{Bauer}, F.~E., {Alexander}, D.~M., {Brandt}, W.~N., {Schneider}, D.~P.,
  {Treister}, E., {Hornschemeier}, A.~E., \& {Garmire}, G.~P. 2004, \aj, 128,
  2048

\bibitem[{{Belczynski} {et~al.}(2008){Belczynski}, {Kalogera}, {Rasio}, {Taam},
  {Zezas}, {Bulik}, {Maccarone}, \& {Ivanova}}]{belczynski08}
{Belczynski}, K., {Kalogera}, V., {Rasio}, F.~A., {Taam}, R.~E., {Zezas}, A.,
  {Bulik}, T., {Maccarone}, T.~J., \& {Ivanova}, N. 2008, \apjs, 174, 223

\bibitem[{{Blanton} {et~al.}(2001){Blanton}, {Sarazin}, \& {Irwin}}]{blanton01}
{Blanton}, E.~L., {Sarazin}, C.~L., \& {Irwin}, J.~A. 2001, \apj, 552, 106

\bibitem[{{Boggess}(1959)}]{boggess59}
{Boggess}, N.~W. 1959, \pasp, 71, 534

\bibitem[{{Broos} {et~al.}(2010){Broos}, {Townsley}, {Feigelson}, {Getman},
  {Bauer}, \& {Garmire}}]{broos10}
{Broos}, P.~S., {Townsley}, L.~K., {Feigelson}, E.~D., {Getman}, K.~V.,
  {Bauer}, F.~E., \& {Garmire}, G.~P. 2010, \apj, 714, 1582

\bibitem[{{Cash}(1979)}]{cash79}
{Cash}, W. 1979, \apj, 228, 939

\bibitem[{{Clauset} {et~al.}(2007){Clauset}, {Rohilla Shalizi}, \&
  {Newman}}]{clauset07}
{Clauset}, A., {Rohilla Shalizi}, C., \& {Newman}, M.~E.~J. 2007, ArXiv
  e-prints

\bibitem[{{Colbert} {et~al.}(2004){Colbert}, {Heckman}, {Ptak}, {Strickland},
  \& {Weaver}}]{colbert04}
{Colbert}, E.~J.~M., {Heckman}, T.~M., {Ptak}, A.~F., {Strickland}, D.~K., \&
  {Weaver}, K.~A. 2004, \apj, 602, 231

\bibitem[{{Davis}(2001)}]{davis01}
{Davis}, J.~E. 2001, \apj, 562, 575

\bibitem[{{Dickey} \& {Lockman}(1990)}]{dickey90}
{Dickey}, J.~M., \& {Lockman}, F.~J. 1990, \araa, 28, 215

\bibitem[{{Eddington}(1913)}]{eddington13}
{Eddington}, A.~S. 1913, \mnras, 73, 359

\bibitem[{{Fabbiano}(1988)}]{fabbiano88}
{Fabbiano}, G. 1988, \apj, 325, 544

\bibitem[{{Fabbiano} \& {White}(2006)}]{fabbiano06}
{Fabbiano}, G., \& {White}, N.~E. 2006, in Compact stellar X-ray sources, ed.
  {Lewin, W.~H.~G.~\& van der Klis, M.} (University of Chicago Press), 475--506

\bibitem[{{Fabbiano} {et~al.}(2010){Fabbiano}, {Brassington}, {Lentati},
  {Angelini}, {Davies}, {Gallagher}, {Kalogera}, {Kim}, {King}, {Kundu},
  {Pellegrini}, {Richings}, {Trinchieri}, {Zezas}, \& {Zepf}}]{fabbiano10}
{Fabbiano}, G., {et~al.} 2010, \apj, 725, 1824

\bibitem[{{Fragos} {et~al.}(2008){Fragos}, {Kalogera}, {Belczynski},
  {Fabbiano}, {Kim}, {Brassington}, {Angelini}, {Davies}, {Gallagher}, {King},
  {Pellegrini}, {Trinchieri}, {Zepf}, {Kundu}, \& {Zezas}}]{fragos08}
{Fragos}, T., {et~al.} 2008, \apj, 683, 346

\bibitem[{{Freedman} {et~al.}(1994){Freedman}, {Hughes}, {Madore}, {Mould},
  {Lee}, {Stetson}, {Kennicutt}, {Turner}, {Ferrarese}, {Ford}, {Graham},
  {Hill}, {Hoessel}, {Huchra}, \& {Illingworth}}]{freedman94}
{Freedman}, W.~L., {et~al.} 1994, \apj, 427, 628

\bibitem[{{Freeman} {et~al.}(2001){Freeman}, {Doe}, \&
  {Siemiginowska}}]{freeman01}
{Freeman}, P., {Doe}, S., \& {Siemiginowska}, A. 2001, in Society of
  Photo-Optical Instrumentation Engineers (SPIE) Conference Series, Vol. 4477,
  Society of Photo-Optical Instrumentation Engineers (SPIE) Conference Series,
  ed. {J.-L.~Starck \& F.~D.~Murtagh}, 76--87

\bibitem[{{Freeman} {et~al.}(2002){Freeman}, {Kashyap}, {Rosner}, \&
  {Lamb}}]{freeman02}
{Freeman}, P.~E., {Kashyap}, V., {Rosner}, R., \& {Lamb}, D.~Q. 2002, \apjs,
  138, 185

\bibitem[{{Fridriksson} {et~al.}(2008){Fridriksson}, {Homan}, {Lewin}, {Kong},
  \& {Pooley}}]{fridriksson08}
{Fridriksson}, J.~K., {Homan}, J., {Lewin}, W.~H.~G., {Kong}, A.~K.~H., \&
  {Pooley}, D. 2008, \apjs, 177, 465

\bibitem[{{Garmire} {et~al.}(2003){Garmire}, {Bautz}, {Ford}, {Nousek}, \&
  {Ricker}}]{garmire03}
{Garmire}, G.~P., {Bautz}, M.~W., {Ford}, P.~G., {Nousek}, J.~A., \& {Ricker},
  Jr., G.~R. 2003, in Society of Photo-Optical Instrumentation Engineers (SPIE)
  Conference Series, Vol. 4851, Society of Photo-Optical Instrumentation
  Engineers (SPIE) Conference Series, ed. {J.~E.~Truemper \& H.~D.~Tananbaum},
  28--44

\bibitem[{{Gehrels}(1986)}]{gehrels86}
{Gehrels}, N. 1986, \apj, 303, 336

\bibitem[{{Gilfanov}(2004)}]{gilfanov04a}
{Gilfanov}, M. 2004, \mnras, 349, 146

\bibitem[{{Gilfanov} {et~al.}(2004){Gilfanov}, {Grimm}, \&
  {Sunyaev}}]{gilfanov04b}
{Gilfanov}, M., {Grimm}, H., \& {Sunyaev}, R. 2004, \mnras, 351, 1365

\bibitem[{{Gordon} {et~al.}(2004){Gordon}, {P{\'e}rez-Gonz{\'a}lez}, {Misselt},
  {Murphy}, {Bendo}, {Walter}, {Thornley}, {Kennicutt}, {Rieke}, {Engelbracht},
  {Smith}, {Alonso-Herrero}, {Appleton}, {Calzetti}, {Dale}, {Draine},
  {Frayer}, {Helou}, {Hinz}, {Hines}, {Kelly}, {Morrison}, {Muzerolle},
  {Regan}, {Stansberry}, {Stolovy}, {Storrie-Lombardi}, {Su}, \&
  {Young}}]{gordon04}
{Gordon}, K.~D., {et~al.} 2004, \apjs, 154, 215

\bibitem[{{Grimm} {et~al.}(2002){Grimm}, {Gilfanov}, \& {Sunyaev}}]{grimm02}
{Grimm}, H., {Gilfanov}, M., \& {Sunyaev}, R. 2002, \aap, 391, 923

\bibitem[{{Grimm} {et~al.}(2003){Grimm}, {Gilfanov}, \& {Sunyaev}}]{grimm03}
---. 2003, \mnras, 339, 793

\bibitem[{{Immler} \& {Wang}(2001)}]{immler01}
{Immler}, S., \& {Wang}, Q.~D. 2001, \apj, 554, 202

\bibitem[{{Johnston} \& {Verbunt}(1996)}]{johnston96}
{Johnston}, H.~M., \& {Verbunt}, F. 1996, \aap, 312, 80

\bibitem[{{Kenter} \& {Murray}(2003)}]{kenter03}
{Kenter}, A.~T., \& {Murray}, S.~S. 2003, \apj, 584, 1016

\bibitem[{{Kilgard} {et~al.}(2002){Kilgard}, {Kaaret}, {Krauss}, {Prestwich},
  {Raley}, \& {Zezas}}]{kilgard02}
{Kilgard}, R.~E., {Kaaret}, P., {Krauss}, M.~I., {Prestwich}, A.~H., {Raley},
  M.~T., \& {Zezas}, A. 2002, \apj, 573, 138

\bibitem[{{Kim} \& {Fabbiano}(2004)}]{kim04b}
{Kim}, D., \& {Fabbiano}, G. 2004, \apj, 611, 846

\bibitem[{{Kim} \& {Fabbiano}(2010)}]{kim10}
---. 2010, \apj, 721, 1523

\bibitem[{{Kim} {et~al.}(2004){Kim}, {Cameron}, {Drake}, {Evans}, {Freeman},
  {Gaetz}, {Ghosh}, {Green}, {Harnden}, {Karovska}, {Kashyap}, {Maksym},
  {Ratzlaff}, {Schlegel}, {Silverman}, {Tananbaum}, {Vikhlinin}, {Wilkes}, \&
  {Grimes}}]{kim04a}
{Kim}, D., {et~al.} 2004, \apjs, 150, 19

\bibitem[{{Kim} {et~al.}(2006){Kim}, {Fabbiano}, {Kalogera}, {King},
  {Pellegrini}, {Trinchieri}, {Zepf}, {Zezas}, {Angelini}, {Davies}, \&
  {Gallagher}}]{kim06}
---. 2006, \apj, 652, 1090

\bibitem[{{Kim} {et~al.}(2007){Kim}, {Wilkes}, {Kim}, {Green}, {Barkhouse},
  {Lee}, {Silverman}, \& {Tananbaum}}]{kim07}
{Kim}, M., {Wilkes}, B.~J., {Kim}, D., {Green}, P.~J., {Barkhouse}, W.~A.,
  {Lee}, M.~G., {Silverman}, J.~D., \& {Tananbaum}, H.~D. 2007, \apj, 659, 29

\bibitem[{{Kolmogorov}(1941)}]{kolmogorov41}
{Kolmogorov}, A. 1941, {Ann. Math. Statist.}, 12, 461

\bibitem[{{Kong} {et~al.}(2003){Kong}, {DiStefano}, {Garcia}, \&
  {Greiner}}]{kong03}
{Kong}, A.~K.~H., {DiStefano}, R., {Garcia}, M.~R., \& {Greiner}, J. 2003,
  \apj, 585, 298

\bibitem[{{Kormendy} {et~al.}(2010){Kormendy}, {Drory}, {Bender}, \&
  {Cornell}}]{kormendy10}
{Kormendy}, J., {Drory}, N., {Bender}, R., \& {Cornell}, M.~E. 2010, \apj, 723,
  54

\bibitem[{{Kotulla} {et~al.}(2009){Kotulla}, {Fritze}, {Weilbacher}, \&
  {Anders}}]{kotulla09}
{Kotulla}, R., {Fritze}, U., {Weilbacher}, P., \& {Anders}, P. 2009, \mnras,
  396, 462

\bibitem[{{Kraft} {et~al.}(1991){Kraft}, {Burrows}, \& {Nousek}}]{kraft91}
{Kraft}, R.~P., {Burrows}, D.~N., \& {Nousek}, J.~A. 1991, \apj, 374, 344

\bibitem[{{Kruskal}(1952)}]{kruskal52}
{Kruskal}, W.~H. 1952, {Ann. Math. Statist.}, 23, 525

\bibitem[{{Kundu} {et~al.}(2002){Kundu}, {Maccarone}, \& {Zepf}}]{kundu02}
{Kundu}, A., {Maccarone}, T.~J., \& {Zepf}, S.~E. 2002, \apjl, 574, L5

\bibitem[{{Liu} {et~al.}(2006){Liu}, {van Paradijs}, \& {van den
  Heuvel}}]{liu06}
{Liu}, Q.~Z., {van Paradijs}, J., \& {van den Heuvel}, E.~P.~J. 2006, \aap,
  455, 1165

\bibitem[{{Maccarone} {et~al.}(2003){Maccarone}, {Kundu}, \&
  {Zepf}}]{maccarone03}
{Maccarone}, T.~J., {Kundu}, A., \& {Zepf}, S.~E. 2003, \apj, 586, 814

\bibitem[{{Mann} \& {Whitney}(1947)}]{mann47}
{Mann}, H.~B., \& {Whitney}, D.~R. 1947, {Ann. Math. Statist.}, 18, 50

\bibitem[{{Meurs} \& {van den Heuvel}(1989)}]{meurs89}
{Meurs}, E.~J.~A., \& {van den Heuvel}, E.~P.~J. 1989, \aap, 226, 88

\bibitem[{{Pence} {et~al.}(2001){Pence}, {Snowden}, {Mukai}, \&
  {Kuntz}}]{pence01}
{Pence}, W.~D., {Snowden}, S.~L., {Mukai}, K., \& {Kuntz}, K.~D. 2001, \apj,
  561, 189

\bibitem[{{Plucinsky} {et~al.}(2008){Plucinsky}, {Williams}, {Long}, {Gaetz},
  {Sasaki}, {Pietsch}, {T{\"u}llmann}, {Smith}, {Blair}, {Helfand}, {Hughes},
  {Winkler}, {de Avillez}, {Bianchi}, {Breitschwerdt}, {Edgar}, {Ghavamian},
  {Grindlay}, {Haberl}, {Kirshner}, {Kuntz}, {Mazeh}, {Pannuti}, {Shporer}, \&
  {Thilker}}]{plucinsky08}
{Plucinsky}, P.~P., {et~al.} 2008, \apjs, 174, 366

\bibitem[{{Prestwich} {et~al.}(2003){Prestwich}, {Irwin}, {Kilgard}, {Krauss},
  {Zezas}, {Primini}, {Kaaret}, \& {Boroson}}]{prestwich03}
{Prestwich}, A.~H., {Irwin}, J.~A., {Kilgard}, R.~E., {Krauss}, M.~I., {Zezas},
  A., {Primini}, F., {Kaaret}, P., \& {Boroson}, B. 2003, \apj, 595, 719

\bibitem[{{Primini} {et~al.}(1993){Primini}, {Forman}, \& {Jones}}]{primini93}
{Primini}, F.~A., {Forman}, W., \& {Jones}, C. 1993, \apj, 410, 615

\bibitem[{{Rosati} {et~al.}(2002){Rosati}, {Tozzi}, {Giacconi}, {Gilli},
  {Hasinger}, {Kewley}, {Mainieri}, {Nonino}, {Norman}, {Szokoly}, {Wang},
  {Zirm}, {Bergeron}, {Borgani}, {Gilmozzi}, {Grogin}, {Koekemoer}, {Schreier},
  \& {Zheng}}]{rosati02}
{Rosati}, P., {et~al.} 2002, \apj, 566, 667

\bibitem[{{Sarazin} {et~al.}(2000){Sarazin}, {Irwin}, \& {Bregman}}]{sarazin00}
{Sarazin}, C.~L., {Irwin}, J.~A., \& {Bregman}, J.~N. 2000, \apjl, 544, L101

\bibitem[{{Shaw Greening} {et~al.}(2009){Shaw Greening}, {Barnard}, {Kolb},
  {Tonkin}, \& {Osborne}}]{shawgreening09}
{Shaw Greening}, L., {Barnard}, R., {Kolb}, U., {Tonkin}, C., \& {Osborne},
  J.~P. 2009, \aap, 495, 733

\bibitem[{{Siegel}(1956)}]{siegel56}
{Siegel}, S. 1956, Nonparametric Statistics for the Behavioral Sciences
  (McGraw-Hill)

\bibitem[{{Soria} \& {Kong}(2002)}]{soria02b}
{Soria}, R., \& {Kong}, A.~K.~H. 2002, \apjl, 572, L33

\bibitem[{{Soria} \& {Wu}(2002)}]{soria02a}
{Soria}, R., \& {Wu}, K. 2002, \aap, 384, 99

\bibitem[{{Supper} {et~al.}(1997){Supper}, {Hasinger}, {Pietsch}, {Truemper},
  {Jain}, {Magnier}, {Lewin}, \& {van Paradijs}}]{supper97}
{Supper}, R., {Hasinger}, G., {Pietsch}, W., {Truemper}, J., {Jain}, A.,
  {Magnier}, E.~A., {Lewin}, W.~H.~G., \& {van Paradijs}, J. 1997, \aap, 317,
  328

\bibitem[{{Swartz} {et~al.}(2003){Swartz}, {Ghosh}, {McCollough}, {Pannuti},
  {Tennant}, \& {Wu}}]{swartz03}
{Swartz}, D.~A., {Ghosh}, K.~K., {McCollough}, M.~L., {Pannuti}, T.~G.,
  {Tennant}, A.~F., \& {Wu}, K. 2003, \apjs, 144, 213

\bibitem[{{Tennant} {et~al.}(2001){Tennant}, {Wu}, {Ghosh}, {Kolodziejczak}, \&
  {Swartz}}]{tennant01}
{Tennant}, A.~F., {Wu}, K., {Ghosh}, K.~K., {Kolodziejczak}, J.~J., \&
  {Swartz}, D.~A. 2001, \apjl, 549, L43

\bibitem[{{Trinchieri} \& {Fabbiano}(1991)}]{trinchieri91}
{Trinchieri}, G., \& {Fabbiano}, G. 1991, \apj, 382, 82

\bibitem[{{Trinchieri} {et~al.}(1999){Trinchieri}, {Israel}, {Chiappetti},
  {Belloni}, {Stella}, {Primini}, {Fabbiano}, \& {Pietsch}}]{trinchieri99}
{Trinchieri}, G., {Israel}, G.~L., {Chiappetti}, L., {Belloni}, T., {Stella},
  L., {Primini}, F., {Fabbiano}, P., \& {Pietsch}, W. 1999, \aap, 348, 43

\bibitem[{{Trudolyubov} {et~al.}(2002){Trudolyubov}, {Borozdin}, {Priedhorsky},
  {Mason}, \& {Cordova}}]{trudolyubov02}
{Trudolyubov}, S.~P., {Borozdin}, K.~N., {Priedhorsky}, W.~C., {Mason}, K.~O.,
  \& {Cordova}, F.~A. 2002, \apjl, 571, L17

\bibitem[{{Voss} {et~al.}(2009){Voss}, {Gilfanov}, {Sivakoff}, {Kraft},
  {Jord{\'a}n}, {Raychaudhury}, {Birkinshaw}, {Brassington}, {Croston},
  {Evans}, {Forman}, {Hardcastle}, {Harris}, {Jones}, {Juett}, {Murray},
  {Sarazin}, {Woodley}, \& {Worrall}}]{voss09}
{Voss}, R., {et~al.} 2009, \apj, 701, 471

\bibitem[{{White} \& {Ghosh}(1998)}]{white98}
{White}, N.~E., \& {Ghosh}, P. 1998, \apjl, 504, L31+

\bibitem[{{Williams} {et~al.}(2008){Williams}, {Gaetz}, {Haberl}, {Pietsch},
  {Shporer}, {Ghavamian}, {Plucinsky}, {Mazeh}, {Sasaki}, \&
  {Pannuti}}]{williams08}
{Williams}, B.~F., {et~al.} 2008, \apj, 680, 1120

\bibitem[{{Wise} {et~al.}(2003){Wise}, {Davis}, {Huenemoerder}, {Houck}, \&
  {Dewey}}]{wise03}
{Wise}, M.~W., {Davis}, J.~E., {Huenemoerder}, D.~P., {Houck}, J.~C., \&
  {Dewey}, D. 2003, {MARX 4.0 Technical Manual}, (Cambridge: MIT ),
  ftp://space.mit.edu/pub/CXC/MARX/v4.0/marx\_4.0\_manual.pdf

\bibitem[{{Wu} {et~al.}(2003){Wu}, {Tennant}, {Swartz}, {Ghosh}, \&
  {Hunstead}}]{wu03}
{Wu}, K., {Tennant}, A.~F., {Swartz}, D.~A., {Ghosh}, K.~K., \& {Hunstead},
  R.~W. 2003, ArXiv Astrophysics e-prints

\bibitem[{{Zezas} \& {Fabbiano}(2002)}]{zezas02b}
{Zezas}, A., \& {Fabbiano}, G. 2002, \apj, 577, 726

\bibitem[{{Zezas} {et~al.}(2006){Zezas}, {Fabbiano}, {Baldi}, {Schweizer},
  {King}, {Ponman}, \& {Rots}}]{zezas06}
{Zezas}, A., {Fabbiano}, G., {Baldi}, A., {Schweizer}, F., {King}, A.~R.,
  {Ponman}, T.~J., \& {Rots}, A.~H. 2006, \apjs, 166, 211

\bibitem[{{Zezas} {et~al.}(2007){Zezas}, {Fabbiano}, {Baldi}, {Schweizer},
  {King}, {Rots}, \& {Ponman}}]{zezas07}
{Zezas}, A., {Fabbiano}, G., {Baldi}, A., {Schweizer}, F., {King}, A.~R.,
  {Rots}, A.~H., \& {Ponman}, T.~J. 2007, \apj, 661, 135

\bibitem[{{Zezas} {et~al.}(2002){Zezas}, {Fabbiano}, {Rots}, \&
  {Murray}}]{zezas02a}
{Zezas}, A., {Fabbiano}, G., {Rots}, A.~H., \& {Murray}, S.~S. 2002, \apj, 577,
  710

\end{thebibliography}

\end{document}